%% file: main.tex
\documentclass[sigconf,nonacm]{acmart}
\AtBeginDocument{%
  }

\setcopyright{none}
\copyrightyear{}
\acmYear{}
\acmDOI{}
\acmISBN{}
\acmConference[]{}{}{}
\renewcommand\footnotetextcopyrightpermission[1]{}
\makeatletter
\renewcommand\@formatdoi[1]{}
\renewcommand\@copyrightpermission{}
\makeatother

\usepackage{xspace}
\usepackage{xcolor}
\usepackage{ifthen}
\usepackage{multirow}
\usepackage{booktabs}
\usepackage{tabularx}
\usepackage{longtable}
\usepackage{subcaption} %
\usepackage{listings}
\usepackage[normalem]{ulem} %
\begin{document}

\title{Mimetic Alignment with ASPECT: Evaluation of AI-inferred Personal Profiles}

\author{Ruoxi Shang}
\email{rxshang@uw.edu}
\affiliation{%
  \institution{University of Washington}
  \city{Seattle}
  \state{Washington}
  \country{USA}
}

\author{Dan Marshall}
\email{Dan.Marshall@microsoft.com}
\affiliation{%
  \institution{Microsoft Research}
  \city{Redmond}
  \state{Washington}
  \country{USA}
}

\author{Edward Cutrell}
\email{cutrell@microsoft.com}
\affiliation{%
  \institution{Microsoft Research}
  \city{Redmond}
  \state{Washington}
  \country{USA}
}

\author{Denae Ford}
\email{denae@microsoft.com}
\affiliation{%
  \institution{Microsoft Research}
  \city{Redmond}
  \state{Washington}
  \country{USA}
}

\renewcommand{\shortauthors}{Shang, et al.}

\begin{abstract}
\input{sections/01-abstract}

\end{abstract}

\ccsdesc[500]{Human-centered computing~Collaborative and social computing systems and tools}
\ccsdesc[500]{Human-centered computing~Empirical studies in collaborative and social computing}
\ccsdesc[300]{Human-centered computing~User studies}
\keywords{communication style, psychometric profiling, LLM personalization, AI agents, mimetic alignment, user auditing}

\received{26 March 2026}

\maketitle

\input{sections/00-macro}

\input{sections/02-introduction_v4}

\input{sections/03-related-work.tex}

\input{sections/05-pipeline-1-profiling_v2}

\input{sections/05-pipeline-2-audit_v2}

\input{sections/06-study-design}

\input{sections/08-findings-RQ1.tex}

\input{sections/08-findings-RQ2.tex}

\input{sections/08-findings-RQ3}
\input{sections/09-discussion.tex}

\input{sections/limitations}

\input{sections/10-conclusion.tex}

\bibliographystyle{ACM-Reference-Format}
\bibliography{references}

\appendix

\input{sections/appendix}

\end{document}

%% file: sections/01-abstract.tex
AI agents that communicate on behalf of individuals need to capture how each person actually communicates, yet current approaches either require costly per-person fine-tuning, produce generic outputs from shallow persona descriptions, or optimize preferences without modeling communication style.
We present ASPECT (Automated Social Psychometric Evaluation of Communication Traits), a pipeline that directs LLMs to assess constructs from a validated communication scale against behavioral evidence from workplace data, without per-person training.
In a case study with 20 participants (1,840 paired item ratings, 600 scenario evaluations), ASPECT-generated profiles achieved moderate alignment with self-assessments, and ASPECT-generated responses were preferred over generic and self-report baselines on aggregate, with substantial variation across individuals and scenarios.
During the profile review phase, linked evidence helped participants identify mischaracterizations, recalibrate their own self-ratings, and negotiate context-appropriate representations.
We discuss implications for building inspectable, individually scoped communication profiles that let individuals control how agents represent them at work.

%% file: sections/00-macro.tex
\newcommand{\condA}{\textit{Generic}\xspace}
\newcommand{\condB}{\textit{Self-Report}\xspace}
\newcommand{\condC}{\textit{Profiled}\xspace}
\newcommand{\system}{ASPECT\xspace}

\newcommand{\qq}[1]{``\emph{#1}''}

\newcommand{\RQA}{How accurately can LLMs infer individuals' communication styles from workplace data?\xspace}
\newcommand{\RQB}{How effectively do these inferred profiles translate to appropriate social representations?\xspace}
\newcommand{\RQC}{What factors explain variance in both inference accuracy and representation quality? \xspace}

%% file: sections/02-introduction_v4.tex
\section{Introduction}

AI agents are increasingly expected to communicate not just \textit{for} people but \textit{as} them: auto-replying to messages, attending meetings on a user's behalf, or acting as conversational proxies when someone is unavailable~\cite{cheng2025conversational, leong2024dittos, patel2024zoom}.
For these agents to be useful, they must capture how a specific individual communicates, not produce a generic approximation~\cite{lee2023speculating, huang2025mirror}.
Unlike public figures with documented personas that models can learn from~\cite{characterai2025}, ordinary individuals have private, sparse, and context-dependent communication patterns.
A person may be direct with peers and diplomatic with leadership, verbose in brainstorming and terse in status updates.
Representing someone faithfully means capturing these patterns from their actual behavior.
To date, existing approaches to personalization fall short.
Persona prompting, where the model is given a short description of who to imitate, produces outputs that look superficially different but are similar across individuals~\cite{zakazov2024assessing, cheng2024gems}.
Per-person fine-tuning or preference-based alignment (RLHF, DPO) can capture individual patterns but requires substantial data and compute for each person, does not scale, and produces opaque models that the individual cannot inspect or adjust.

We introduce \textbf{\system{} (Automated Social Psychometric Evaluation of Communication Traits)}, a pipeline that uses LLMs guided by a validated psychometric scale to extract representative snippets from a person's social interaction data and build a structured, evidence-grounded communication profile without per-person training. As each dimension score is linked to specific quotes and rationales, the profile is interpretable—the person can inspect where it is aligned or not.
Whether such a profile is socially appropriate, however, can only be determined by the person it represents~\cite{hwang2024whose, cheng2025conversational, hancock2020ai}.
We therefore conducted a case study with 20 participants from a single organization to assess what the pipeline captures and where it falls short (Section~\ref{sec:auditing}).
Each participant first reviewed their inferred profile against their own self-ratings, inspecting the behavioral evidence and rationales behind each construct.
They then evaluated scenario-based responses generated under three conditions: \condC{} (from \system{}-inferred profile), \condB{} (based upon self-ratings only), and \condA{} (vanilla GPT with no personalization).

Our evaluation reveals three key patterns: profiled responses were preferred over both baselines on aggregate, though with substantial individual and scenario variation; LLMs exhibited a consistent positivity bias, over-rating socially desirable traits; and profile review prompted participants to recalibrate their own self-ratings and negotiate context-appropriate representations.
\noindent This paper contributes:
\begin{enumerate}
    \item \textbf{\system{}}, a profiling pipeline that operationalizes a validated psychometric instrument as a prompt-based data compression method, generating structured communication profiles that are portable across models and inspectable by the people they describe.
    \item \textbf{Empirical evidence} from 20 participants providing the first systematic account of LLM-based communication profiling evaluated by the individuals being modeled, revealing that inference preserves relative profile shape but introduces predictable biases—findings that directly inform calibration strategies for future systems.
    \item \textbf{Design implications} for human-in-the-loop profile negotiation: we demonstrate that evidence-linked review shifts profiling from a classification task to a collaborative negotiation, where individuals decide not just whether a profile is accurate but which version of themselves they want an agent to represent.
\end{enumerate}

%% file: sections/03-related-work.tex
\section{Related Work}

\subsection{From communication support to AI representation}

AI-mediated communication (AI-MC) systems help users communicate more effectively by modifying, augmenting, or generating messages for interpersonal objectives \cite{hancock2020ai}. Applications include rewriting for empathy \cite{sharma2021towards}, reframing negative perspectives \cite{ziems2022inducing}, smart reply suggestions \cite{hohenstein2018ai}, persuasion enhancement \cite{jakesch2019ai}, conveying status \cite{pavlick2016empirical} or trustworthiness \cite{ma2017self}, promoting democratic discourse \cite{argyle2023leveraging}, and increasing affectionate communication between partners \cite{kim2019love}. In each case, the user remains the communicator; the AI operates on the message, not on the person's behalf.

A different line of work builds agents that communicate \textit{as} individuals. Dittos are personalized embodied agents that attend meetings when users are unavailable \cite{leong2024dittos}. Recent work has experimented with generative agents that simulate human behavior in virtual environments \cite{park2023generative} and mimetic models predict specific individuals' responses \cite{mcilroy2022mimetic}. Platforms like Character.AI generate representations of public figures and fictional characters \cite{characterai2025}. Public figures have documented personas that models can learn from. For ordinary individuals with private footprints, building an accurate representation remains an open problem, and these developments raise concerns about identity fragmentation, autonomy, and whether individuals can recognize and approve how AI speaks for them \cite{lee2023speculating}.

The dominant paradigm for aligning model behavior to users is preference-based optimization. RLHF \cite{ouyang2022training} and DPO \cite{rafailov2023direct} train models to produce outputs that humans prefer, and recent personalized variants learn per-user reward functions from heterogeneous feedback \cite{poddar2024personalizing, chakraborty2024maxmin}. These methods address a different problem than ours. Preference alignment optimizes which output a user likes; style representation captures how a specific person communicates. A person might prefer concise AI responses while writing lengthy, discursive emails themselves. Preference-based methods also require hundreds to thousands of comparisons per individual, each encoding a binary signal along an unspecified axis, and produce opaque reward functions tied to specific model weights. Representing someone's communication style for social contexts requires a structured, interpretable profile that the person can inspect and adjust. Closest works to ours are agent simulations initialized from survey data reproduce behaviors for over 1,000 individuals \cite{park2024generative}, and general user models infer structured beliefs from everyday computer use traces \cite{shaikh2025creating}. However, these aforementioned approaches do not typically use observational data and experiment with profiling for social communication contexts as we do in our approach.

\subsection{Building structured profiles of individuals}
\label{sec:related-work-2}

Design personas compress user segments into qualitative narratives that ground coordination across teams \cite{nielsen2014personas}. Psychometric instruments take a complementary approach: they decompose individual behavior into quantitative, theory-backed constructs that can be reliably measured and compared across people \cite{cronbach1955construct, devellis2021scale}. The Big Five personality model, for example, captures individual differences across five broad domains and finer-grained facet-level traits using standardized inventories \cite{goldberg1993structure, costa2008revised}. For communication specifically, the Communication Styles Inventory (CSI) models six domains with 23 facets, each operationalized by behaviorally anchored items rated on Likert scales, with validated internal consistency and convergent evidence \cite{devries2013communication}. These instruments provide interpretable dimensions with established reliability, making them candidates for structuring an individual's profile. With foundation models, personality prompting has been used to steer LLM behavior toward specific trait configurations. Recent tests show, however, that personality-prompted LLMs often fail to behave consistently with the intended persona in social decision paradigms \cite{zakazov2024assessing}. A structured profile does not guarantee appropriate downstream behavior; the person being represented needs to verify it.

Behavioral data offers a complementary path to building individual profiles. Digital-footprint inference translates social media traces into Big Five predictions that rival close acquaintances' assessments \cite{kosinski2013private, youyou2015computer}, and data-grounded methods reduce biases inherent in self-report \cite{kosinski2015facebook}. These systems demonstrate that behavioral traces carry stable personality signals, but they output broad trait scores without linking them to specific behavioral evidence that the person can review. General user models take a different approach, accumulating calibrated, revisable beliefs about specific individuals from everyday computer use to drive context-aware assistants \cite{shaikh2025creating}. They model what a person does across applications but do not organize that knowledge around validated communication constructs. At the population scale, agent simulations initialized from qualitative interviews can reproduce survey responses and behaviors for over 1,000 individuals \cite{park2024generative}, showing that it is possible to represent many distinct people rather than a single average user, though their profiles are derived from interview summaries rather than observed interaction data. Domain-specific work has shown the value of adapting communication style for particular contexts, from culturally tailored persuasive agents \cite{zhou2017adapting} to health chatbots adapted for multicultural caregivers \cite{baik2025adapting}, but these adapt style at the group level rather than building individual profiles.

Across this landscape, a gap remains: no existing approach builds structured communication profiles for individuals that are organized by validated psychometric constructs, grounded in behavioral evidence from the person's own data, and open to the person's inspection and adjustment. We explore this direction with \system, which uses a psychometric scale as scaffolding to compress an individual's workplace communication data into a structured, evidence-linked profile, and we assess what this approach captures and where it falls short in one workplace setting.

%% file: sections/05-pipeline-1-profiling_v2.tex
\section{System Design}
\label{sec:system-design}

\textbf{\system} constructs a structured communication profile for a given individual from their workplace interaction data using a \textit{psychometric instrument} as a novel prompt-based data compression method.

\begin{figure*}[h]
  \centering
    \includegraphics[width=0.95\textwidth]{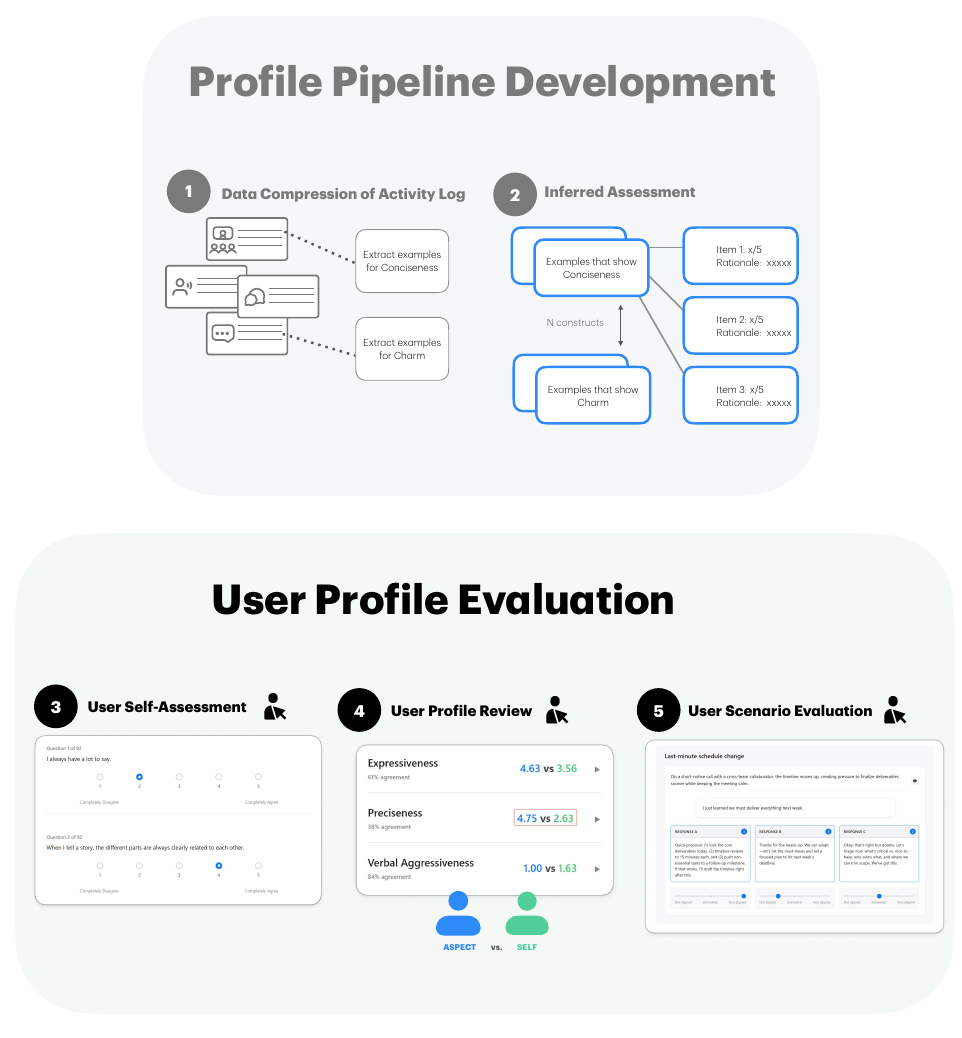}
  \caption{Profile Generation Pipeline and User Evaluation Overview. See text for detailed explanation for each phase}
  \label{fig:ASPECT-overview}
\end{figure*}

\subsection{Conceptual Model}

Language models can simulate individual behaviors and preferences when provided with behavioral data~\cite{zakazov2024assessing, park2024generative, shaikh2025creating}. We build on this capability to construct communication profiles from observed workplace interaction data. While several approaches can steer LLM behavior with personal data, we use prompt-based context engineering for two reasons.

First, it is lightweight. Fine-tuning requires substantial data per person, is costly to maintain, and produces opaque models. Retrieval-augmented generation (RAG) does not fit this task. Retrieval methods would weigh informational relevancy more, and also lead to unstable and inconsistent characteristics across conversations. Second, the approach must be interpretable. As using AI to represent individuals in social communications is still exploratory, there is no consensus on what aspects of communication need to be captured, what works, or why. Computed metrics cannot validate this nuance, and only the individuals themselves can verify whether their profile is socially appropriate. Hence, we need an approach that can render the profile interpretable, allowing people to review each component, identify where it succeeds or fails, and examine probable causes. Given these requirements, we wanted to develop a structured profile prompt with good coverage of communication dimensions. Our challenge was then to determine which dimensions to include and how to extract them from data.

We draw inspiration from psychometric assessment, both as a \textit{method} and as a \textit{measurement instrument} (Section~\ref{sec:related-work-2}). 
As a method, psychometric assessment follows a structured procedure—trained observers review behavioral evidence through specific constructs and score standardized items—and we apply this same logic to LLMs.
Rather than asking the model to characterize a person open-endedly, we direct it to scan communication data construct by construct, collect behavioral evidence for each, and score individual items against the collected evidence. As an instrument, a validated psychometric scale provides the construct scaffolding that organizes this process. A construct such as \textit{Talkativeness} in the CSI~\cite{devries2013communication} is operationalized by behaviorally anchored items, each rated on a Likert scale. Such scales report continuous scores across dimensions and facet-level subtraits rather than assigning a categorical ``type.'' Their validated structure supports internal consistency and stability across contexts.

The resulting profile is a set of construct-level ratings, each linked to behavioral evidence. It aggregates across many interactions rather than reflecting a single event. We instantiate this approach using the CSI; Section~\ref{sec:evaluation-in-workplace} details the specific instrument and data.

\subsection{Profile Pipeline Development}

The profiling pipeline implements this approach in two phases. We used OpenAI's o1 reasoning model for all inference tasks. %

\paragraph{Phase 1: Construct-guided Evidence Extraction.} The social interaction data contains authentic behavior but is vast, sparse, and noisy. The challenge is how to compress it into a structured representation of an individual's communication patterns. The pipeline first preprocesses the user's communication data (e.g., meeting transcripts, group chats) into batches organized by token budget. It then scans the data \textit{facet by facet}: for each of the scale's constructs, the LLM receives one batch of communication files along with the construct definition and its associated scale items. The model is instructed to act as an objective observer and identify 2--5 concrete instances where the user's behavior demonstrates that specific construct. Scanning one facet at a time ensures that the model attends to one behavioral pattern across the full breadth of the data, rather than attempting to characterize the user holistically in a single pass. Each evidence record pairs a context summary covering the situation, social dynamics, setting, and behavioral analysis with a conversational excerpt of 2--5 turns, keeping judgments traceable to source data (Appendix~\ref{app:evidence-schema}).

\paragraph{Phase 2: Inferred Assessment from Item-level Scoring.} Once evidence has been collected for all constructs, the pipeline scores each individual scale item independently. For a given item (e.g., ``I always have a lot to say''), the model receives the item text together with the behavioral evidence extracted for its parent facet in Phase~1. It then produces a numerical rating (1--5) and a brief rationale grounding the rating in the observed evidence. Items within the same facet share the same evidence pool, ensuring scoring consistency within constructs. When no behavioral evidence was found for a facet, the pipeline assigns default ratings that reflect absence of the behavior rather than leaving scores undefined.

\noindent An example item-level output:
\begin{lstlisting}
{"Item": "I sometimes toss bizarre ideas into a group discussion.",
 "Rating": "4/5",
 "Rationale": "Across multiple brainstorming sessions, the user
  repeatedly suggests unconventional ideas (novel workflow
  automations, experimental feature integrations). These proposals
  show a pattern of introducing unusual concepts, aligning well
  with the idea of occasionally tossing unexpected suggestions
  into group discussions."}
\end{lstlisting}

Evidence extraction and scoring address different questions (``what did this person do?'' vs.\ ``how well does this item describe them?''), and decoupling them allows each step to be inspected and validated independently.

%% file: sections/05-pipeline-2-audit_v2.tex
\begin{figure*}[h]
  \centering
  \includegraphics[width=0.9\textwidth]{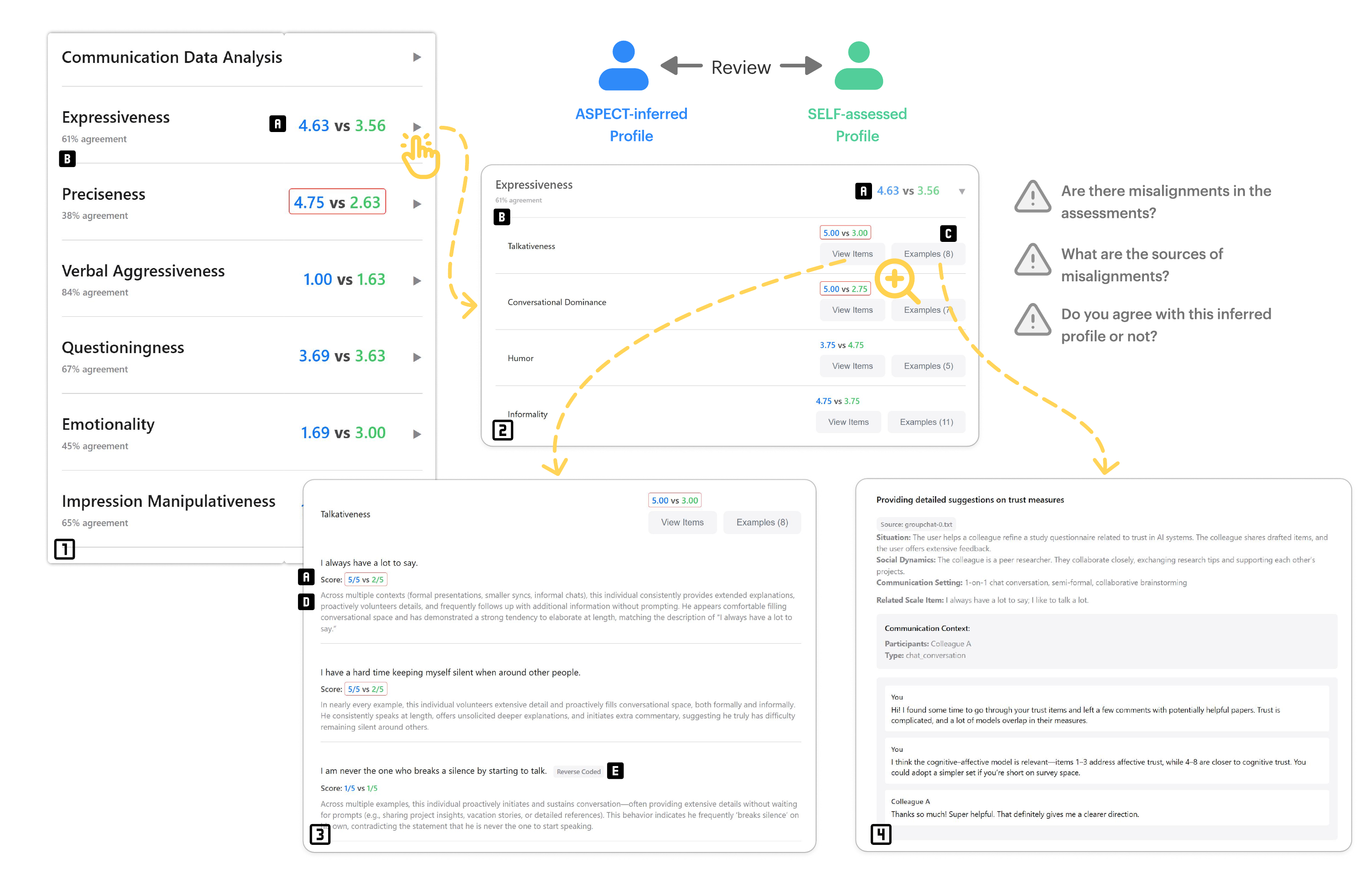}
  \caption{Detailed outline of Phase 4: User Profile Review.}
  \label{fig:review-interface-compare}
\end{figure*}

Our evaluation asks whether the profiling pipeline produces profiles that individuals recognize as capturing their communication style, and whether those profiles generate appropriate AI behavior in context. The first does not guarantee the second, because a person might accept their trait scores yet find AI responses generated from them stilted or situationally off. We therefore evaluate at both the profile level and the behavioral level.
Since representation preferences are personal~\cite{hwang2024whose, cheng2025conversational, hancock2020ai}, participants serve as the primary evaluators at both the profile and behavioral levels.
We use participant not as ground-truth accuracy measure of our system, but to provide insight where the pipeline succeeds, fails, and what alignment means if they were to use this in practice.

\paragraph{Phase 3: User Self-Assessment} To ground the profile review, participants first complete a self-assessment on the same instrument the pipeline scored (the CSI we mentioned previously), providing per-item comparison data. Both stages are implemented as a Flask web application with standard HTML/JavaScript, served locally to each participant. The application guides participants through self-assessment, profile review with side-by-side score comparisons and linked evidence, and blinded scenario evaluation. Participants can flag factual errors or unrepresentative evidence during review, and all responses and preferences are persisted as JSON for analysis.

\subsection{User Profile Evaluation}
\label{sec:auditing}

\paragraph{Phase 4: User Profile Review}
\label{sec:auditing-profile}

The review interface mirrors the instrument's ontological nature (Figure~\ref{fig:review-interface-compare} mirrors Table~\ref{tab:csi_main_text}). At the top level, participants navigate the CSI's dimension-facet-item structure~(1). Expanding a dimension reveals individual items~(2), the statements that both participants and \system rate directly. During self-assessment, participants see only raw items without the ontological context, so they are not primed by construct labels. Each item links to behavioral evidence extracted from the user's data~(3): a representative conversational excerpt of 2--5 turns with the model's rationale.

Several interface elements support comparison and verification as highlighted in Figure~\ref{fig:review-interface-compare}: (A)\system and self-ratings appear side by side, with red highlighting for disagreements of 2 or more points; (B) percent agreement across ratings; (C) an example count per construct shows where data coverage is thin; (D) the system's full rationale is displayed for each score;  and (E) a reverse-coded indicator flags items that measure the opposite direction of their parent construct. Appendix~\ref{app:csi-full} includes additional details.

\paragraph{Phase 5: User Scenario-Based Evaluation}
\label{sec:auditing-behavior}

Scenario-based evaluation tests whether the inferred profile produces appropriate responses in concrete workplace situations. Scenarios must be personally relevant so that participants can meaningfully judge whether a response captures how they would communicate, but they must also be structurally comparable across participants so that responses can be analyzed across the sample. Fully personalized scenarios for each participant would make cross-participant comparison impossible; identical generic scenarios would lack ecological validity.

We separate scenario \textit{structure} from scenario \textit{content}. Structure comes from the APRACE taxonomy (Actor--Partner--Relation--Activities--Context--Evaluation) \cite{hoppler2022aprace}, which decomposes interpersonal situations into independent dimensions that map to workplace communication constructs. We authored 10 scenario templates spanning the six CSI dimensions, each defined by a fixed configuration of eleven APRACE factors such as hierarchy, familiarity, purpose, and stakes. Templates were stratified to balance hierarchy $\times$ purpose $\times$ stakes, ensuring every attribute level appears at least once (Appendix~\ref{appendix:scenarios}). Every participant's Scenario~1 shares the same interpersonal setup: the same power dynamic, relational context, communication purpose, and stakes level. What differs is the specific content.

\begin{figure*}[h]
  \centering
  \includegraphics[width=0.9\textwidth]{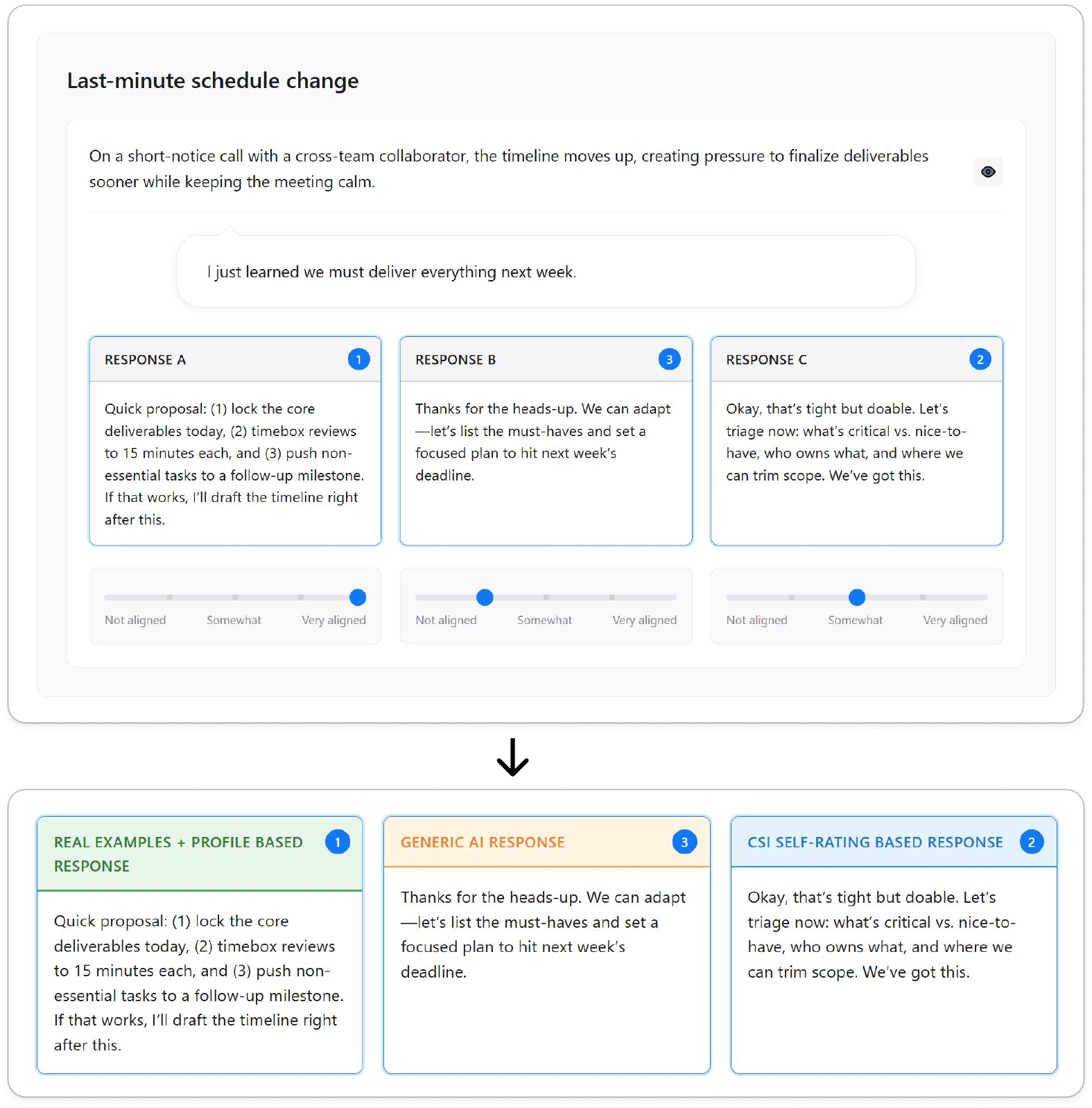}
  \caption{Scenario-based assessment interface}
  \label{fig:audit-interface-scenario}
\end{figure*}

Each template is instantiated with participant-specific details drawn from recent work communication, such as team names, tools, and terminology. In earlier pilot testing, participants struggled to project themselves into generic workplace situations and could not meaningfully assess which response best captured their style; grounding scenarios in familiar details resolved this. Scenarios remain hypothetical (``what if\ldots'') rather than retrospective (``remember when\ldots''), so participants judge potential behavior, not recall. Every participant sees structurally equivalent scenarios set in their own workplace context (Figure~\ref{fig:audit-interface-scenario}). Across participants, the content varies but the interpersonal setup stays fixed.

For each scenario, three responses are generated under different conditions and presented in randomized order without labels. Participants rank the responses and rate each on a 5-point alignment scale (1 = not aligned, 5 = very aligned), then may choose to reveal which condition produced each. The conditions vary in how much personal information the model receives. \condA\ is a scenario-only baseline where the model sees only the scenario and partner message. \condB\ adds a conversational style description derived from the participant's self-ratings but no behavioral examples. \condC\ uses a style description derived from the \system profile and includes compact behavioral-evidence snippets to ground the response.

The three conditions form a controlled comparison. Both \condB\ and \condC\ convert their respective CSI ratings into the same style-description format using the same prompt; the difference is whether those ratings come from the participant or from the pipeline's analysis of their data, isolating the rating source while holding instrument, format, and conversion constant. 
\condB\ uses structured CSI ratings rather than free-form descriptions, so any difference in response quality traces to rating source alone, not input format.
\condC\ also includes behavioral evidence because evidence extraction is how the pipeline arrives at its ratings: the model scores each item against behavioral instances collected from the person's data, so evidence and ratings are not separable components. All three conditions receive the same scenario and partner message, and system role prompts are matched to minimize confounds.

Ideally, scenario-based evaluation would follow one or more rounds of profile review so that downstream behavior is tested on a refined profile. Because the feasibility of our profiling method was unknown, we treat this as a first-step feasibility study and condition scenario responses on the initial, pre-review profile.

%% file: sections/06-study-design.tex
\section{Study Design}

\subsection{Evaluation in a Workplace Setting}
\label{sec:evaluation-in-workplace}
We evaluate in a workplace setting using communication traces that participants can access through enterprise meeting and chat systems. To capture natural, unstructured  communication in everyday professional relationships, each session participant exported the prior 90 days of their meeting transcripts, group messages, and direct message chat histories and processed them locally on their own devices. This was a deliberate choice motivated by a few factors. Organizations routinely retain transcripts and chat logs, offering a relatively easy way for participants to retrieve their own data. Working within a single organization also introduces shared norms of tone, tools, and schedules. Conversations are situated in real teams, tools, and deadlines, yet share organizational norms. This reduces cross-domain noise and supports more controlled comparisons. Next, workplace traces avoid many intimate topics that arise in family and friend communications and are thus easier to audit with minimal intrusion. Finally, we wanted participants to retain control over their personal data.

Broad personality batteries (e.g., Big Five) capture stable dispositions yet remain one step removed from conversational behavior, and team instruments (e.g., FIRO-B, DiSC, TKI) are often vendor-specific and scenario-bound. To model how people actually communicate, we use the Communication Styles Inventory (CSI)~\cite{devries2013communication}, a behaviorally anchored measure developed from a lexical study of communication descriptors. CSI contains 92 items\footnote{The original CSI numbers items 1--96; we omit the Inscrutableness facet (4 items) due to poor psychometric properties (see Appendix~\ref{app:csi-full}).} distributed across six domain-level scales—\emph{Expressiveness}, \emph{Preciseness}, \emph{Verbal Aggressiveness}, \emph{Questioningness}, \emph{Emotionality}, and \emph{Impression Manipulativeness}—each split into four four-item facets, yielding interpretable axes directly tied to observable speech behavior rather than intrapersonal cognitions. 
In the original validation, all domain reliabilities exceeded $\alpha{=}0.80$, factor analyses supported the six-domain, 24-facet structure in a large community sample ($N{=}815$) with replication in a student sample, and the scales showed medium-to-high convergent validity with lexical marker scales and behavior-oriented communication measures while remaining discriminant from nonbehavioral ``about-communication'' cognitions~\cite{devries2013communication}. CSI also shows coherent relations with personality (medium-to-strong associations with HEXACO and NEO dimensions), and related work demonstrates incremental validity of communication styles for leadership criteria beyond extraversion and conscientiousness, underscoring that styles add predictive signal beyond broad traits~\cite{bakker2013incremental}. These properties make CSI a theory-driven, validated, and interpretable scaffold for compressing workplace discourse into facet-level evidence that we can audit transparently later in our pipeline. 

The study protocol was reviewed and approved by the organization's research ethics board.
\input{sections/tables/csi_main_text}

\subsection{Participants}
We recruited $N{=}20$ participants (11 men, 9 women) within a single large organization, aged 18–54, spanning interns to senior-level staff across research, engineering, and program/project management roles. Tenure at the organization ranged from $<$6 months to $>$10 years; time-in-role ranged from $<$3 months to $>$3 years. Work arrangements varied across fully in-office ($n{=}9$), hybrid ($n{=}10$), and fully remote ($n{=}1$). Recruitment was conducted via internal email lists and group channels to intentionally sample diverse roles and collaboration contexts. Participants were compensated \$100 for a single 2-hour session.

\subsection{Procedure}
Each session lasted two hours and was conducted one-on-one via enterprise videoconferencing. In the first $\sim$30 minutes, participants installed the study application on their local laptop, processed their exported workplace data locally for profiling, and completed the 92-item CSI self-assessment. 

The remaining $\sim$90 minutes followed the two auditing processes outlined in Section~\ref{sec:auditing}: in \textit{profile auditing}, participants reviewed model scores with linked behavioral evidence and refined interpretations; in \textit{behavior auditing}, they evaluated personalized scenario responses by rating alignment on a 1–5 scale and ranking three anonymized responses generated from different personalization sources, with the within-scenario order randomized. All collection and processing occurred locally on participant devices, including a locally hosted interface for the self-ratings, profile auditing, and ranking and rating for the scenarios.

During the self-assessment, participants completed a 92-item version of the CSI scale. Items were presented in the instrument’s original interleaved questionnaire order (facets dispersed across the 1–96 list rather than blocked by facet) to limit response-set bias, and rated on a 1–5 Likert scale. See Appendix~\ref{app:csi-full} for item order. CSI is a behaviorally anchored measure validated across diverse settings~\cite{devries2013communication}, so constraining ratings to workplace contexts is a standard contextualization aligned with our downstream workplace scenario evaluation. Participants were specifically instructed to judge their work-self only—i.e., to base each rating on how they typically communicate in professional settings and to exclude non-work contexts.

%% file: sections/tables/csi_main_text.tex
\begin{table*}[h]
\centering
\small
\begin{tabular}{p{2.5cm}p{2.5cm}p{7cm}}
\toprule
\textbf{Dimension} & \textbf{Facet} & \textbf{Example Item} \\
\midrule
\multirow{4}{*}{\textbf{Expressiveness}} 
    & Talkativeness & ``I like to talk a lot.'' \\
    & Conv. Dominance & ``I often take the lead in a conversation.'' \\
    & Humor & ``My jokes always draw a lot of attention.'' \\
    & Informality & ``I address others in a very casual way.'' \\
\midrule
\multirow{4}{*}{\textbf{Preciseness}} 
    & Structuredness & ``My stories always contain a logical structure.'' \\
    & Thoughtfulness & ``I think carefully before I say something.'' \\
    & Substantiveness & ``Conversations with me always involve some important topic.'' \\
    & Conciseness & ``I don't need a lot of words to get my message across.'' \\
\midrule
\multirow{4}{*}{\parbox{2.5cm}{\textbf{Verbal\\Aggressiveness}}} 
    & Angriness & ``I tend to snap at people when I get annoyed.'' \\
    & Authoritarianism & ``I expect people to obey when I ask them to do something.'' \\
    & Derogatoriness & ``I have at times made people look like fools.'' \\
    & Nonsupportiveness & ``I always show a lot of understanding for other people's problems.'' (R) \\
\bottomrule
\end{tabular}
\vspace{0.5em}
\caption{A subset of the CSI scale we used for this system. The full CSI contains 6 dimensions, 23 facets, and 92 items rated on a 5-point scale (1=strongly disagree, 5=strongly agree). (R) indicates reverse-coded items. See Appendix \ref{app:csi-full} for full scale.}
\label{tab:csi_main_text}
\end{table*}

%% file: sections/08-findings-RQ1.tex
\section{Findings}
Participants generally described \system-generated profile to be meaningful and insightful, largely because the extracted evidence felt concrete and on-point. Participants consistently commented: \qq{The examples are good… it did a good job} (P2); \qq{I really love how it took the data and the examples… for each category} (P6); \qq{these examples are actually really good} and the social-dynamics analysis was  \qq{very interesting} (P12). P12 also noted the system correctly surfaced multiple humor instances unique to them. Others explicitly accepted the assessments: \qq{I accept [it]… that’s a correct assessment} (P14). P10 highlighted, 
\qq{
    I think that its assessment of my personality is better than my own assessment of my personality, at least in a professional setting and that is delightful.
    }

Beyond surface-level linguistic cues, the system recovered latent patterns and habits. P10 noted it correctly inferred a dispositional tendency to ``run meetings'' even when not leading. P18 and P6 realized the ``structured'' assessments and examples correctly reflected their consistent practice of pre-polishing messages before sending them out, even though they considered themselves naturally unstructured; P13 made this explicit, \qq{I try very hard to be structured… drafted… before sending} and affirmed it generalizes to new-group interactions \qq{This is how I talk}.

To understand how \system works beyond these impressions, we conducted a systematic mixed method evaluation for our entire pipeline design through two evaluation axes: the accuracy of the inferred profile and its downstream utility for response generation. The following sections (Section~\ref{sec:findings-rq1} and Section~\ref{sec:findings-rq2}) will be organized by two research questions. 

\textbf{\emph{RQ1 Inference alignment} - How accurately can an LLM infer individuals’ communication-style profiles from workplace data?} Compared to the self-report of over 92 items ($N{=}20$; $1{,}840$ pairs), alignment is meaningful but imperfect ($\mathrm{MAE}{=}1.39$, weighted $\kappa{=}0.34$, within-person mean $\rho{=}0.39$; dimension-level $\rho$ up to $0.72$). 
We analyzed the review process as an \emph{analytic probe and calibration} of this initial inference: in $\sim17.7\%$ of facet reviews, participants either adopted the AI score or negotiated a middle ground, indicating that discrepancies often reflect self-bias, construct interpretation, or coverage limits rather than pure model error. 

\textbf{\emph{RQ2 Behavioral alignment} - Is social imitation via a \system\ profile \emph{sufficient} to produce responses that participants judge as appropriate in real workplace scenarios?} 
In a blinded, within-subject triad (600 evaluations), profile-conditioned responses win $42.5\%$ of first-place ranks (vs.\ $32.5\%$ Generic, $25.0\%$ Self-Report), receive higher mean alignment ratings ($3.33$ vs.\ $3.09$ vs.\ $2.95$), and are preferred over Self-Report (Friedman $\chi^{2}{=}9.31$, $p{=}.0095$; Wilcoxon $p{=}.0067$) with a small but significant lift over Generic (LMM $\beta{=}0.24$, $p{=}.045$). Preferences depend on person and scenario type.

\subsection{RQ1: Inference Alignment}
\label{sec:findings-rq1}
\input{sections/tables/table_all_sidebyside}

\subsubsection{Statistical Results}
\label{sec:rq1-quant}
\paragraph{What we measured and why.}
To test whether the profiling stage provides a good \emph{starting profile} (Evaluation frame), we compared \system’s ratings to participants’ self-ratings. We analyze $N{=}20$ participants with 1{,}840 paired item ratings ($92{\times}20$), aggregated to 23 facets and 6 dimensions. All self-ratings were anchored to the \emph{work-self} and administered in the instrument’s interleaved order to reduce response sets (Section~\ref{sec:evaluation-in-workplace}).

We report: (i) \emph{exact match} and \emph{mean absolute error} (MAE) to summarize numeric closeness on the 1–5 scale; (ii) \emph{bias} (signed mean difference) to detect systematic over-/under-estimation by \system; (iii) \emph{agreement beyond chance} using weighted $\kappa$; (iv) \emph{rank correlations} $\rho$ to capture profile \emph{shape} (relative highs/lows) either within-person (across items for each participant) or between-people (for a given trait, does \system order participants like self-reports?); and (v) \emph{ICCs} [ICC(A,1) for absolute agreement; ICC(C,1) for consistency] to summarize two-rater reliability. Items are treated as integers; facets and dimensions are arithmetic means of their items.

\paragraph{Overall alignment.}
Across items, exact numeric matches occur in $23.8\%$ of cases; MAE is $1.39$ on the 1–5 scale (95\% CI $[1.34,1.45]$), indicating typical disagreements of about one category. Agreement beyond chance is fair (weighted $\kappa=0.34$). Importantly, within-person rank correlation averages $\rho=0.39$ (95\% CI $[0.31,0.44]$): even when absolute values differ, \system often recovers each person’s relative highs and lows (Table~\ref{tab:overall}).

\paragraph{Dimension-level patterns.}
Alignment is heterogeneous by dimension. Verbal Aggressiveness and Emotionality show the smallest errors and fair reliability (e.g., Verbal Aggressiveness: MAE $=0.59$, bias $=-0.40$, between-person $\rho=0.39$, ICC(A,1) $=0.35$; Emotionality: MAE $=0.74$, bias $=-0.05$, $\rho=0.20$, ICC(A,1) $=0.31$), suggesting \system captures overt interpersonal tone reasonably well. In contrast, Expressiveness (MAE $=1.03$, bias $=+0.99$, $\rho=0.48$) and especially Preciseness (MAE $=1.69$, bias $=+1.69$, $\rho=0.18$, ICCs $\approx 0$) show larger numeric gaps with over-rating by \system, indicating a calibration need for structural/organizational traits.

\paragraph{Facet-level patterns.}
Facets with clear linguistic signals align better—for example, Angriness (MAE $=0.51$, bias $=-0.49$, $\rho=0.34$, ICC(A,1) $=0.25$), Charm (MAE $=0.65$, bias $=-0.08$, ICC(A,1) $=0.26$), and Humor (MAE $=0.81$, bias $=-0.11$, $\rho=0.43$, ICC(A,1) $=0.33$). Facets that hinge on subtle intent or context, such as Nonsupportiveness (MAE $=0.89$, bias $=-0.49$, low ICCs) and Inquisitiveness (MAE $=0.90$, bias $=+0.48$, ICCs $\approx 0$), are more challenging. Bland–Altman limits typically span about $\pm2$ points across facets, reflecting genuine between-person variability.

\paragraph{Item-level examples.}
High agreement concentrates on directly observable behavior (e.g., ``Even when I’m angry, I won’t take it out on someone else.''), where MAE runs $\sim 0.25$–$0.50$ and exact matches can exceed $60\%$. Lowest agreement arises for abstract tendencies (e.g., “Conversations with me always involve some important matter.”), where MAE can exceed $2$ and $\kappa \approx 0$.

\paragraph{Response-style check.}
To rule out scale-use artifacts (e.g., some people avoid 1s/5s), we z-standardized ratings within participant for both self and \system and recomputed correlations. Results are unchanged to three decimals at all levels (items: mean $\rho=0.386$ raw vs.\ $0.386$ standardized; facets: $0.466$ vs.\ $0.466$; dimensions: $0.720$ vs.\ $0.720$), indicating that discrepancies largely reflect substantive differences rather than response styles. Cosine similarities (items $=0.384$, facets $=0.468$, dimensions $=0.732$) tell the same story.

\paragraph{Within- vs.\ between-participant agreement.}
Within individuals, \system preserves profile shape (median $\rho$ in Table~\ref{tab:overall}: items $=0.46$, facets $=0.55$, dimensions $=0.72$). Between participants, \system better differentiates who is more \emph{Expressive} (dimension-level $\rho=0.48$) and moderately distinguishes \emph{Emotionality} ($\rho=0.20$). Between-person discrimination at finer levels is weak (facet/item $\rho$ often near zero), which is expected given limited observations per facet.

\paragraph{Reliability context.}
Where underlying measures are internally consistent, agreement improves. Several self facets have high $\alpha$ ($\geq 0.80$) while \system’s reliabilities vary (e.g., strong on Conversational Dominance, weaker on Talkativeness). Averaged across items, ICC(A,1) $=0.345$ and ICC(C,1) $=0.349$ indicate fair two-rater reliability, with the strongest dimension-level ICCs for Verbal Aggressiveness (up to $0.45$) and the weakest for Preciseness ($\sim 0$), mirroring MAE/bias patterns.

\paragraph{Summary.} 
In this sample, \system recovered recognizable communication patterns for most participants. Absolute alignment is modest (MAE $\approx 1.4$, $\kappa \approx 0.34$), but relative agreement is stronger, especially at the dimension level,
suggesting the pipeline preserves the relative shape of individuals' profiles even when absolute scores diverge. Systematic biases (e.g., over-rating on Preciseness) signal where calibration can help. As we show next, the auditing stage leverages these signals: participants review evidence, negotiate definitions, and adjust scores, turning the initial profile into a user-aligned representation suited for downstream scenarios.

\subsubsection{Auditing as Bidirectional Alignment}
\label{sec:bidirectional-alignment}

Through profile auditing, participants compared side-by-side scores, read evidence-linked rationales, clarified how constructs were defined, recalled additional context not captured in the examples, and assessed whether the pipeline behaved consistently. We find that quantitative misalignment in Section~\ref{sec:rq1-quant} were often more nuanced than error alone, and auditing operates as a calibration and reflection step rather than a simple accept/reject check.

Participants examined \system's outputs for each of the 23 facets against their self-assessments while thinking aloud and reviewing ratings, linked examples, and rationales. This yielded 411 facet-level evaluations (20 participants $\times$ 23 facets, minus cases without explicit feedback). Three researchers independently coded a subset, reconciled differences over three rounds, and one researcher verified all instances before thematic analysis. The category counts below are descriptive indicators of the overall distribution. These numbers should be interpreted as approximate rather than exact. We estimate that on the order of 5--10 cases per category could be borderline or mislabeled despite reconciliation. Each audit decision was coded into mutually exclusive categories:
\begin{itemize}
  \item \textit{Totally Aligned} ($n{=}141$): Clear agreement; evidence accepted.
  \item \textit{Misalign--Disapprove} ($n{=}142$): Participants defended the self-rating and rejected the model’s.
  \item \textit{Misalign--Middle Ground} ($n{=}42$): Participants deemed both sides partly right and sought a compromise.
  \item \textit{Misalign--Approve AI} ($n{=}31$): Participants revised their self-rating to match the model after reviewing evidence.
  \item \textit{Unsure/Not Mentioned} ($n{=}55$): No clear decision or insufficient rationale (often due to limited evidence).
\end{itemize}

\paragraph{Interpretation.}
Through this process, we find auditing both \emph{reduces resolvable error} and \emph{documents principled, context-dependent differences}. Participants \emph{changed their initial self-ratings} in $\sim17.7\%$ ($73/411$) facet reviews: $\sim7.5\%$ ($31/411$) fully adopted the AI score and  $\sim10.2\%$ ($42/411$) negotiated an explicit middle ground. In our coding process, we noticed a nontrivial share of ``misalignments'' reflected individuals' self-rating bias. This indicates auditing produces measurable movement toward evidence when accurately provided. This also means that the pre-audit itemwise statistical metrics (e.g., MAE, $\kappa$ in Section~\ref{sec:rq1-quant}) are conservative lower bounds on alignment.  We conducted thematic analysis across these five different audit decisions, and the following sections detail common recurring themes.

\subsubsection{Where \system matched participants' self-assessments}

Direct, unambiguous alignments happened frequently when it was assessing stable or salient communication styles of participants. For example, P20 on talkativeness: \qq{I think it aligns with me pretty well. I always have a lot to say… I like to explain and elaborate things… which are correct.} P14 on humor: \qq{AI and I see face to face.} When \system surfaced structured chains (overviews, bullets, ordered steps), participants confirmed these as accurate habits. P20: \qq{Consistently provides clear sequence… bullet points… This is what I would love to do actually.} P10: \qq{I like that it thinks that my stories always have logical structure.} Participants who habitually ask questions or float unconventional takes recognized themselves immediately. P10 on unconventionality: \qq{[I have] a very high rate of hot takes per minute, and the AI seems to be capturing that.} P13 on inquisitiveness: \qq{I matched the AI on this one.} P13 on charm: \qq{This one is actually pretty accurate… This is how I talk, especially with a new group of people.}

We also find a portion of exact matches when the system finds \emph{no behavioral examples} for a facet and, by design, instantiated a low (negative) endorsement. For example, for an item under \emph{Angriness}, ``I tend to snap at people when I get annoyed.,'' if no evidence is retrieved the item is set to 1 (Strongly Disagree), reflecting a low level of \emph{Angriness}. The pattern was clear for from the distribution in Figure~\ref{fig:no_examples} in Appendix that these no-evidence cases cluster around negative traits in the constructs. Take \emph{Angriness} as an example: P2 noted, \qq{It rated me one. I rated myself 1.75… I do believe I’ve never been angry in the workplace}, and P17 echoed, \qq{I feel like that is very reflective of me}. P1 also commented that all the facets under \emph{Impression Manipulativeness} are \qq{largely in line} by noting there was \qq{not as much data for these}.

\subsubsection{Correct inference and self-rating bias emerge upon reflection}

In many cases, the examples reminded participants of behaviors they had not fully considered. P13, when shown examples of tense interactions, said \qq{this actually happened and I actually felt really uneasy at the time,} acknowledging that they had initially underestimated their tension. P10 also expressed surprise at the system’s evidence of humor, saying \qq{I said that I don’t often make other people laugh and it says actually no, you do… thanks AI.} P4 described underestimating their own talkativeness and concluded that \qq{the AI was right and I was maybe… underestimating how much I speak.} These cases show how surfacing forgotten or minimized evidence led participants to adjust their self-view.

Through reflection, participants often realized they had been too critical of themselves. After reviewing examples of their structured communication, P15 noted \qq{maybe I’m just a little harsh on myself}. The same participant reconsidered their ratings on conciseness and tension, concluding they had been more capable than they had acknowledged. P16 also recognized that their structured way of \qq{spelling out the steps} was clearer than they had credited themselves for and said they \qq{would bump myself up.}

Participants noticed a discrepancy between how they see themselves or wish to be seen as versus how they present themselves to others. Because participants sometimes self rate by referring to an implicit ``ideal self'', while \system can observe only the latter (the enacted, audience-shaped behavior in workplace traces), several ``misalignments'' led to interesting self reflection. P13 noted that being talkative in meetings did not imply comfort: \qq{it doesn’t necessarily mean I’m very comfortable when I talk a lot.} They also emphasized that their structured style was effortful and professional rather than natural: \qq{This is… not really who I am. It’s about how I want to be presented.} P10 similarly recognized a gap between aspiration and observation: they preferred to appear less sentimental or worried at work, yet the AI correctly identified higher levels of sentimentality and worry in their communications. For P19, conversational dominance was behaviorally accurate but role-driven: \qq{part of my job is to direct the conversation,} not a stable personal style.

\subsubsection{Calibrations are needed to find middle ground}

Twenty-four percent of facet evaluations (99/411) resulted in participants seeking a middle ground between self and AI ratings. P4 made this explicit: \qq{I should have rated myself a point higher and it should have rated me a point lower. And then we would have agreed.} These negotiations occurred through mechanisms supported by information provided in the auditing process, when participants could specify exactly what made each assessment partially correct. 

Participants recognized that their behavior varied systematically across contexts and some of the misalignments come from that. P13 on talkativeness: \qq{The workplace situation would prompt me to be very talkative during meetings. It doesn't necessarily mean I'm very comfortable.} They acknowledged the AI correctly captured their meeting behavior (supporting a high rating) while their self-assessment reflected discomfort with that behavior (supporting a lower rating). They eventually decided on a 4/5 compromise that encode both contexts. P19 demonstrated the same pattern with unconventionality: \qq{I do [toy with wild ideas]... but I don't do it necessarily in conversation.}

When confronted with mixed behavioral evidence, some participants explicitly calculated averages. P15 on conciseness: \qq{Is this like 75 examples of short and sustained versus one or two examples of very lengthy? Therefore, four out of five.} They weren't confused—-they were computing a frequency-weighted score. P11 made the same calculation for talkativeness, noting \qq{many of the time I also respond with a yes or a done,} leading to 4/5 rather than AI's 5/5. P17 similarly reasoned through multiple facets by weighing contradictory evidence: extensive speaking in some meetings, silence in others.

Middle ground also emerged when participants discovered mismatches in measurement definitions. P12 initially rated themselves 1/5 on ingratiation, interpreting it as sycophancy. After seeing examples of routine compliments, they did not fully accept the AI's 4/5 but chose 2-3 to reflect both their new understanding and their moderate use of compliments. This pattern repeated across constructs with loaded terms. P15 accepted being humorous but rejected \qq{teasing} as a characterization. P19 agreed they used charm but disputed \qq{flirting.} Participants corrected their understanding while maintaining boundaries around unwanted labels.

\subsubsection{Sources of mischaracterizations and errors}

\input{sections/tables/rationales-misalignment}

To clearly lay out and compare the types of mischaracterizations together, we presented those in Table~\ref{tab:misalignment-taxonomy}, summarizing sources of these issues and reasons participants gave when self and model scores diverged.

%% file: sections/tables/table_all_sidebyside.tex
\begin{table*}[t]
\centering
\small
\caption{Agreement Summary: Overall, Dimensions, and Facets (Side-by-side)}
\begin{minipage}[t]{0.48\linewidth}
\centering
\textbf{Overall and Dimensions}\\
\begin{tabular}{lcc}
\toprule
Metric & Value & 95\% CI\\
\midrule
Exact Match \% & 23.8 & --\\
MAE (absolute difference) & 1.39 & [1.34, 1.45]\\
Weighted Kappa & 0.34 & --\\
Mean Within-Person $\rho$ & 0.39 & [0.31, 0.44]\\
\bottomrule
\end{tabular}
\vspace{4pt}\\
\begin{tabular}{lccc}
\toprule
Dimension & MAE & Bias & $\rho$\\
\midrule
Verbal Aggressiveness & 0.59 & -0.40 & 0.39\\
Emotionality & 0.74 & -0.05 & 0.20\\
Questioningness & 0.74 & 0.23 & 0.26\\
Impression Manipulativeness & 0.82 & -0.47 & -0.03\\
Expressiveness & 1.02 & 0.99 & 0.48\\
Preciseness & 1.69 & 1.69 & 0.18\\
\bottomrule
\end{tabular}
\end{minipage}\hfill%
\begin{minipage}[t]{0.48\linewidth}
\centering
\textbf{Facets (by $\rho$)}\\
\begin{tabular}{lccc}
\toprule
\multicolumn{4}{l}{\emph{Highest agreement}}\\
Facet & MAE & Bias & $\rho$\\
\midrule
Humor & 0.81 & -0.11 & 0.43\\
Sentimentality & 1.29 & 0.79 & 0.39\\
Defensiveness & 1.80 & -1.80 & 0.38\\
Angriness & 0.51 & -0.49 & 0.34\\
Worrisomeness & 1.54 & 0.81 & 0.26\\
\addlinespace[2pt]
\midrule
\multicolumn{4}{l}{\emph{Lowest agreement}}\\
\midrule
Thoughtfulness & 1.27 & 1.27 & -0.26\\
Informality & 1.51 & 1.51 & -0.21\\
Derogatoriness & 0.91 & -0.34 & -0.17\\
Structuredness & 1.39 & 1.39 & -0.16\\
Inquisitiveness & 0.90 & 0.47 & -0.15\\
\bottomrule
\end{tabular}
\end{minipage}
\label{tab:overall}
\end{table*}

%% file: sections/tables/rationales-misalignment.tex
\begin{table*}[t]
\centering
\setlength{\tabcolsep}{5pt}
\renewcommand{\arraystretch}{1.2}
\footnotesize
\begin{tabularx}{\textwidth}{@{}p{2.5cm} X X p{4.0cm}@{}}
\toprule
\textbf{Theme} & \textbf{Concise definition} & \textbf{Typical signals} & \textbf{Representative exemplar} \\
\midrule
\textbf{T1. Coverage \& observability gaps} \textit{Data limitation} &
Ratings inferred from a narrow slice of digital/recorded data; misses offline talk, pre-record small talk, multimodal cues, \emph{internal states}, and rare-but-salient incidents. &
Channel/recording limits; observer effects; offline/in-person only; internalized affect; low-frequency events not visible. &
``We do 5 minutes of informal talk before recording; AI only sees recorded formal content.'' [P15] 
``Philosophical discussions happen offline (in person), not on Teams.'' [P5] 
``Comments affect me emotionally, but I don’t show it in written communication.'' [P11]\\
\addlinespace
\textbf{T2. Situational norms misread as traits} \textit{Method issue} &
Role-, task-, or meeting-driven behavior (e.g., PM leading, presentations, risk triage) inferred as stable dispositional style. &
PM/presenter talk required; task-necessitated structure/precision; professional risk language \(\neq\) personal anxiety; discretion \(\neq\) concealingness. &
``Presentations require me to talk; that doesn’t mean I’m generally talkative.'' [P2] 
``‘Worry’ here reflects PM risk management, not personal anxiety.'' [P15] 
``Swap in any teammate in these meetings and they’d also look dominant.'' [P16] \\
\addlinespace
\textbf{T3. Tone \& valence misinterpretation} \textit{Fundamental LLM limitation} (amplified by data limits) &
Pragmatics are literalized: sarcasm, playful banter, emojis, and praise are reinterpreted as aggression, anxiety, manipulation, or command-giving. &
Sarcasm literalized; joking directives read as authoritarian; emoji/politeness read as stress; praise/cheerleading \(\neq\) ingratiation/charm. &
``The ‘dirty green walls’ line was sarcasm, not sentimentality.'' [P14]
``The ‘Don’t do it’ interruption was playful banter with close colleagues.'' [P19]
``Praise for interns was genuine appreciation, not a manipulation tactic.'' [P15]
\\
\addlinespace
\textbf{T4. Construct / item misalignment} \textit{Method issue} &
Participant and instrument use different definitions or item scopes (operationalization drift; ambiguous boundaries). &
Constructive challenge \(\neq\) argumentative; technical theory \(\neq\) philosophicalness; item asks about tears vs. AI coding general emotion; politeness \(\neq\) sentimentality. &
``Constructive challenge $\neq$ argumentative.'' [P1]
``Theory/framework talk is technical, not ‘philosophical.’'' [P2] 
``The item asks about tears; the AI coded general emotion instead.'' [P12] \\
\addlinespace
\textbf{T5. Evidence use \& scoring integrity problems}\ \textit{Method + tooling issue} &
Overreach from single examples; cherry-picking brief messages; contradictory facet pairings; speaker misattribution; contamination; inconsistent ``no evidence \(\rightarrow\) mid score'' logic. &
Single-example inflation; brevity cherry-pick; logical contradictions; mic-host attribution; third-party input contamination. &
``One worry example shouldn’t make all worry items 5/5.'' [P2] 
``The AI rated me both concise and talks-a-lot—contradictory.'' [P11] 
``As meeting host, others’ comments were attributed to me.'' [P4] \\
\bottomrule
\end{tabularx}
\caption{Typology of sources of AI–self misalignment (RQ1), grouped by root cause: \emph{Data limitations} (T1), \emph{Method issues} (T2, T4, T5), and \emph{Fundamental LLM limitations} (T3). Themes are not mutually exclusive; multi-coding is allowed.}
\label{tab:misalignment-taxonomy}
\end{table*}

%% file: sections/08-findings-RQ2.tex
\subsection{RQ2: From Profiles to Social Performance}
\label{sec:findings-rq2}
While \system demonstrates meaningful inference of communication styles (Section~4.1), accurate trait assessment does not automatically translate to appropriate social representation. We therefore examined whether inferred profiles generated contextually appropriate responses across 10 workplace scenarios, comparing three response conditions: \condA, \condB, and \condC. 

\subsubsection{Statistical Results}
\input{sections/tables/response_preference}
Across 600 evaluations (20 participants $\times$ 10 scenarios $\times$ 3 conditions), responses generated from \condC were preferred over the alternatives on aggregate.

\paragraph{Win rates.} \condC~ responses were ranked first in 42.5\% of scenarios (85/200, 95\% CI [35.9\%, 49.4\%]), compared to 32.5\% for \condA~ (65/200) and 25.0\% for \condB~ (50/200, 95\% CI [19.5\%, 31.4\%]). 

\paragraph{Rankings.} Mean ranks followed the same hierarchy: \condC~ ($M=1.84, SD=0.82$), \condA~ ($M=2.00, SD=0.81$), \condB~ ($M=2.15, SD=0.79$). To test whether these differences were systematic, we conducted a Friedman test (a non-parametric alternative to repeated-measures ANOVA appropriate for ordinal ranks). Results showed a significant main effect of condition ($\chi^2=9.31$, $p=.0095$, Kendall's $W=0.023$). Post-hoc Wilcoxon signed-rank tests with Holm–Bonferroni correction were used to compare conditions pairwise. Only the comparison between \condC~ and \condB~ remained significant ($p=.0067$, $r=.22$), while differences between \condC~ and \condA~, and between \condA~ and Self, did not reach significance.

\paragraph{Ratings.} On the 5-point alignment scale, \condC~ again led ($M=3.33, SD=1.29$), followed by \condA~ ($M=3.09, SD=1.28$) and \condB~ ($M=2.95, SD=1.19$). Because ratings are continuous and approximately interval-scaled, we analyzed them using linear mixed-effects models with random intercepts for participant and scenario. The simplified model (due to convergence limits) revealed that \condC~ were rated significantly higher than \condA~ responses ($\beta=0.24$, $p=.045$), while \condB~ did not significantly differ ($\beta=-0.14$, $p=.26$). Pairwise Cohen's $d$ confirmed small but consistent effects: $d=0.30$ for \condC~ vs. \condB, $d=0.19$ for \condC vs. \condA~.

\paragraph{Summary.} On aggregate, the data suggest a preference ordering \textbf{\condC~ $>$ \condA~ $>$ \condB}. Participants’ own self-reported profiles performed worst, even below \condA~ baselines. This suggests that CSI ratings alone, without behavioral grounding, do not translate into more aligned responses; adding behavioral evidence from communication history improves alignment within our sample.

\subsubsection{Individual Variation Dominates Aggregate Patterns}

Despite aggregate preferences for \condC-generated responses, individual differences were substantial. Random slopes analysis revealed wide variability in condition effects (SD = 0.87 for \condB~, SD = 0.94 for \condC), exceeding the conventional 0.5 threshold for meaningful heterogeneity. 

Nine participants (45\%) showed strong preferences for \condC~, two (10\%) strongly preferred \condB~, and the remaining nine (45\%) exhibited mixed or weak preferences. For example, Participant~2 consistently rated \condC~responses over a point higher than \condA~ on the 5-point scale, whereas Participant~19 showed the opposite, preferring \condB~ by nearly a point. 

Agreement across participants was extremely low (Kendall's $W=0.077$), highlighting that individuals showed low concordance on what constituted a good response. What one participant rated as perfectly aligned ($5/5$), another might rate as misaligned ($2/5$). Scenario-specific patterns were also observed (e.g., \condC~ dominated Scenario~1, whereas \condA~ responses were favored in Scenario~9), though systematic scenario-type effects require further analysis.

On aggregate, \textbf{\condC responses were preferred over \condB~and \condA~ responses, but individual variations dominated the pattern}. However, the strong individual differences observed here underscore the need for adaptive personalization, allowing users to calibrate how much the system should rely on behavioral inference versus self-report.

\subsubsection{Signals of Condition-based Preferences}

\begin{figure*}[h]
  \centering
  \includegraphics[width=0.8\textwidth]{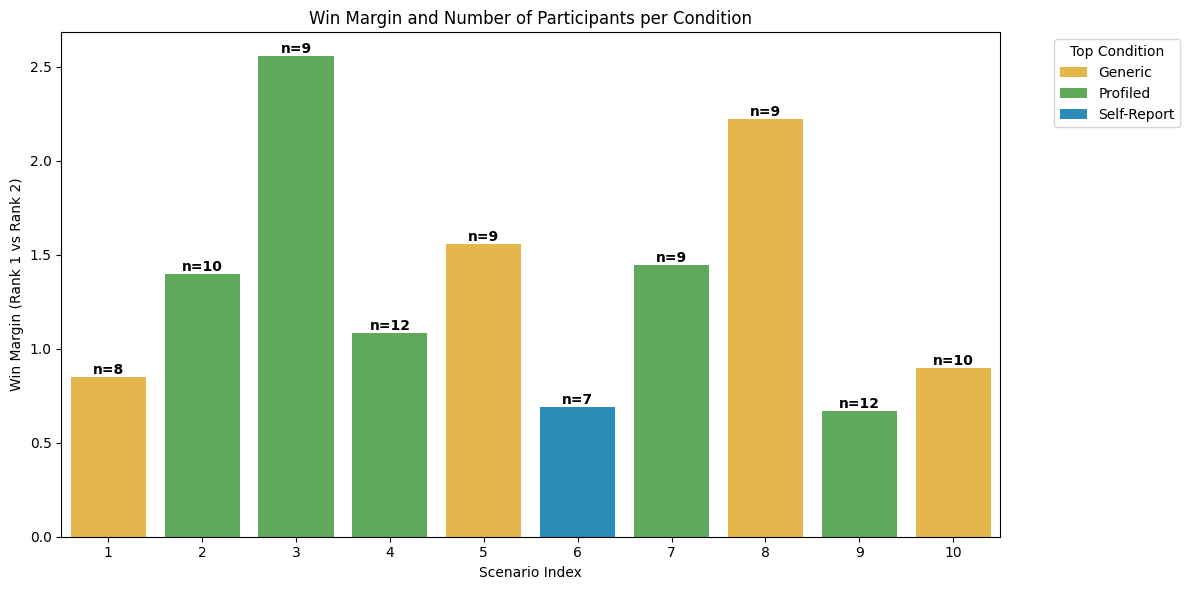} 
  \caption{Win margin of each condition across participants and 10 scenarios.}
  \label{fig:win_margin}
\end{figure*}

Figure \ref{fig:win_margin} shows, for each of ten scenarios, which condition won and how strong the agreement was. Bar color is the winning condition (\condA, \condB, or \condC) chosen by the majority of participants. 
Bar height is the win margin, defined as the number of participants who ranked the winner first minus the number who ranked the runner-up first. The label ``n='' on each bar is how many of the 20 participants selected the winner. 
Across 10 scenarios, \textbf{\condC} wins in 5, \textbf{\condA} wins in 4, and \textbf{\condB} wins in 1. 
Large margins indicate strong agreement (e.g., S3 and S8); small margins (e.g., S9) indicate mixed preferences despite a clear majority.

We cross-analyzed each scenario’s winner and margin against its APRACE metadata (purpose, hierarchy, familiarity, stakes, formality, and mode) to identify patterns between scenario type and preferred condition. We focus on the preference for \condA~and \condC since it is difficult to see pattern given \condB has only one win and the margin is minimal. Participants tend to prefer \textbf{\condC} in scenarios when the task is tightly defined and success depends on the usual way of organizing and speaking. This includes explaining their work to a distant peer (S1), responding in chat when a peer questions their approach (S3), planning a new initiative with a manager in a formal setting (S4), giving a concise standup update (S7), and discussing credit attribution with a distant peer (S9). These settings reward structure, pacing, and the balance between firmness and support that show up in a person’s real communication history. On the other hand, participants tend to prefer \textbf{\condA} when the goal is light coordination or the emotional tone is unclear. This includes setting the tone at a weekly check-in (S1), handling a last-minute schedule change (S5), acknowledging an unexpected process change (S8), and planning a team celebration (S10). This is likely because neutral, generic phrasing is \textit{sufficient} and \textit{safer} in those scenarios. We further analyzed qualitative data of participants' rationales to understand this.

Overall participants showed a general preference for responses that sounded more like themselves or their personal style. However, across the scenarios we noticed patterns of conditions that better captured participants' style.
We noticed that participants who preferred \condC often also referenced how the scenario captured a fuller perspective of how they wanted to be captured, including details on their enthusiasm or approach. For instance, P10 highlights this by indicating that  responses were \qq{very specific about what we would actually do, which I feel like is just...that's how I would talk in a hallway or like that's how I would talk with a co-worker.}
However, this perspective varied across participants. 

In fact, participants who preferred \condA did not like the level of excitement or emotion that may have been out of the norm for how they communicate at work. Specifically, P9 highlighted that \qq{It is trying to be very personal to me, but it is the wrong personal. And so I am deeply uncomfortable with it because I'm like, this isn't my voice.} 
Finally, participants who preferred \condB often cited the low-jargon, more universally collegial way of connecting with colleagues. P14 referred to the \qq{safe} or \qq{corporate  thing to do} when selecting responses. P19 also went on to highlight how these responses were more aligned with their role as a leader. 
These divergent preferences suggest that no single personalization strategy fits all users; effective systems must let individuals choose their level of stylistic adaptation.

%% file: sections/tables/response_preference.tex
\begin{table*}[h!]
\centering
\small
\begin{tabular}{l c c c p{5cm}}
\toprule
\textbf{Cluster} & \textbf{N} & \textbf{Mean Win Rate} & \textbf{Mean Rating Margin} & \textbf{Participants} \\ 
\midrule
Prefers Profiled & 10 & 0.60 & 0.73 & P2, P4, P6, P10, P12, P13, P15, P16, P17, P18 \\ 
Prefers Generic     & 4  & 0.57 & 0.75 & P1, P5, P8, P9 \\ 
Prefers Self-Report        & 2  & 0.70 & 0.85 & P14, P19 \\ 
Mixed / No Clear Preference & 4  & \textit{–} & 0.05 & P3, P7, P11, P20 \\ 
\bottomrule
\end{tabular}
\caption{Participant clusters. We clustered participants by comparing which condition (\condA, \condB, \condC) most often “won” on three summary metrics: win rate (proportion of first-place ranks), mean rank, and mean rating. To avoid over-interpreting small differences, we required either (a) a condition to win at least two of the three metrics and show a meaningful margin ($\geq0.20$ in win rate, $\geq0.25$ in rating, or $\geq0.20$ in rank), or (b) to win one metric with a strong margin ($\geq0.30$, $\geq0.40$, $\geq0.30$ respectively). These thresholds were chosen to align with the observed separations in our data (median margins $\approx0.30$ for win rate, 0.55 for rating, 0.40 for rank, all with IQRs spanning 0.1–0.7), ensuring they capture differences larger than typical within-participant variability. Participants without a clear advantage for any condition were labeled as Mixed. Sensitivity checks with slightly stricter or looser cutoffs shifted at most one participant per cluster, suggesting our results are robust.}
\end{table*}

%% file: sections/09-discussion.tex
\section{Discussion}
\label{sec:discussion}

\subsection{What we learned about building data-grounded profiles}
Our design, implementation, and evaluation of \system revealed three factors that shaped profile quality in our study: (1) having data to ground profile to concrete behavioral evidence; (2) a good psychometric scale that fits the person's profiling needs; and (3) keeping users in the loop to add contextual insight into profile data and provide profile evaluations.

\textbf{Data.} In our study, participants reviewed \system's construct ratings by inspecting behavioral evidence 
and short rationales; this grounded design led them to revise or negotiate their self-ratings. Going forward, we can strengthen this process by showing uncertainty and coverage, preserving provenance for each 
example, or keeping versioned audit trails. These directions align with work on documentation and auditing—Model Cards for model 
reporting~\cite{mitchell2019model}, Datasheets for dataset transparency~\cite{gebru2021datasheets}, and internal 
algorithmic auditing frameworks that support appropriate trust calibration~\cite{raji2020closing,lee2004trust}.

\textbf{A good scale.} In this work, we used the Communication Styles Inventory (CSI) to test the \system framework, and our findings show that not all CSI facets are equally observable in workplace text and some invite interpretation mismatches at the facet level. This is precisely where review helped participants clarify. The implication is to support instrument interpretation alignment or even tailoring with alternative communication scales when a construct is ill-fitted for a user or communication channel. Prior validation work on CSI provides the psychometric grounding and also highlights where adaptations make sense; future versions of \system can learn data-driven observability maps (i.e., which facets are detectable for a given data source) and recommend the right instrument per context~\cite{devries2013communication,diotaiuti2020psychometric}.

\textbf{Users must stay in the loop as reviewers.} One finding from our study is that the review workflow: side-by-side comparisons, linked evidence, and explicit user actions to calibrate assessment supported meaningful calibration. Future systems should extend this approach with more features to support both profile and behavior review. For example, the system could implement more advanced ways for counter-evidence retrieval to combat confirmation bias and coverage diagnostics that surface missing data. Moreover, the system could afford user examining the change in alignment by showing instant change in response style across downstream tasks. Automated evaluation could also be implemented based on some user manual review history to help quickly flag problematic cases. These features are consistent with established guidance on algorithmic auditing, documentation, and trust-calibrated human-AI collaboration~\cite{hoppler2022aprace,raji2020closing,mitchell2019model,gebru2021datasheets,lee2004trust}.

\subsection{Representation boundaries and sources of mischaracterization}

Besides the instances where users calibrated their own score, we also found various types of true mischaracterizations (Table~\ref{tab:misalignment-taxonomy}). Interestingly, not all mischaracterizations should be completely removed, as many participants wish their digital representation to be not exactly like them. The specific design of \system maps out two boundaries in the generated profile through data input and the specific choice of psychometric constructs. That means, the generated profile is constructing a person as how they would behave in a socially constructed context, as P14 described as showing the ``local minima/maxima'' of their characteristics. In our study, this context is naturally the work version of self. We find that Participants often defended context-appropriate self-presentation and drew lines between their natural tendency, learned professional behavior, and desired workplace image. In these cases, ``creating a behavioral average'' would actually be undesirable, as they would want to hide away part of self at work. These findings argue for boundary-respecting profiles: constrain by constructs (what to measure) and by data source (what type of behaviors to be profiled). This aligns with classic accounts of impression management~\cite{goffman1959presentation} and politeness/face-work~\cite{brown1987politeness}, as well as audience design and accommodation—people adapt style to addressees and power relations~\cite{bell1984language,giles1991accommodation}. Because our review process made these boundaries visible, future work could explore context-scoped representations: allow users to pin profiles per context and usage, preview responses under those settings, and opt-in to downplay traits they can exhibit but prefer not to signal in a given setting. For example, a ``workplace casual chat'' agent tuned for warmth and humor.

Other divergences in Table~\ref{tab:misalignment-taxonomy} reflect actionable limits in our current system. To address these issues, future work could explore supplementing textual data with other modalities of data sources as well as engineering more validations to model outputs.

\paragraph{Systematic biases as a learned insight.} Our evaluation revealed that the pipeline systematically over-rates certain dimensions, particularly Preciseness (MAE$=$1.69, bias$=$$+$1.69, ICC$\approx$0) and Expressiveness (bias$=$$+$0.99). Two mechanisms explain this pattern. First, when the pipeline finds no behavioral evidence for a construct, it assigns a low endorsement score. This is informative for traits whose absence in workplace text is meaningful (e.g., no evidence of angry outbursts likely indicates low Angriness), but less so for traits that are simply hard to observe in text (e.g., internal deliberation before speaking). Second, workplace communication is filtered through professional norms: messages tend to be edited, structured, and purposeful, which inflates ratings on constructs like Preciseness regardless of the individual's natural tendencies. These biases were not predictable before the study; they emerged through review, as participants flagged constructs where the data systematically overstated their traits. This finding directly informs future calibration work: once bias patterns are identified across a population of users, automated corrections become feasible (e.g., dimension-specific shrinkage or coverage-based weighting). Our study provides an initial empirical basis for designing such corrections in this setting.

\subsection{Using \system framework in practice}

\textbf{Mimetic agents you can review and deploy.}
In our scenarios, \condC responses were preferred over \condA and \condB conditions on aggregate, and participants generally considered the response generated is aligned with their style. This means these profiles are already actionable for context-specific agents.  \system produces a ``portable'' profile that can be used directly as an agent prompt once generated. However, we found that participants exhibited a threshold effect in their tolerance for misalignment. P9, explains this uncanny valley of social representation in detail: \qq{I can wear a generic Halloween face mask that's not molded to my face at all, and that's fine. But if I have something that's molded improperly to my face, it's deeply uncomfortable.} This captures a fundamental challenge for mimetic agents: imperfect personalization often feels worse than no personalization at all. As participants reviewed the scenarios, we noticed the formation of an implicit alignment with interpretation. For instance, participants also mentioned 95\% vs 70\% type of alignment. There is an unspoken threshold of when it could be better to go with the \condA~ option vs. \condC~the profiled option. At times it felt like when it came down to accurate representation without achieving it, the more unsettling the experience became. This suggests that mimetic AI systems must either achieve very high fidelity or clearly signal their limitations to avoid falling into this uncomfortable middle ground where users experience their digital selves as distorted rather than absent. These distortions and working through the evolutions of them is a field of work that should continue to be studied across contexts.

\textbf{Evidence-based profile for self-reflection.} Although we did not intend to design \system\ as a reflection tool, participants repeatedly engaged in reflective practices: they reconsidered their habits when reading linked examples and justifications. Interestingly, though the review involves reading lots of text, most participants found it engaging and interesting to ``learn about themselves''. This opens up opportunities to leverage the \system framework and this type of individual social profiles to support self-reflection practices, aligning with literature on self-explanation effect for deeper understanding~\cite{chi1994eliciting}, and feedback-intervention theory for crafting feedback that improves performance without triggering defensiveness~\cite{kluger1996effects}. 

\textbf{A de-biasing aid for psychometric assessments.} We found that seeing evidence and rationales upon misalignment triggers calibration and reduces bias in self-assessment. This is in line with a large literature showing that self-assessments are noisy and biased, from classic meta-analysis on limited validity~\cite{mabe1982validity} to Dunning–Kruger miscalibration~\cite{kruger1999unskilled}, and to SOKA results and meta-analyses where informants sometimes predict behavior better than the self~\cite{vazire2010knows,connelly2010other}. \system's approach could make psychometric self-assessments more valid by using data inference as an external anchor and observer of an individuals' actions.

%% file: sections/limitations.tex
\section{Limitations}

\paragraph{\textbf{Limited scope.}}
 For practical reasons elaborated in Section~\ref{sec:evaluation-in-workplace}, we tested \system on only one source of data (workplace conversations), one psychometric scale (CSI), and participants from a single organization.  Recruiting within one organization was necessary for trust (participants knew their data stayed within enterprise boundaries) but limits generalizability. Our 20 participants skewed toward technical roles comfortable with AI systems. We cannot claim it generalizes to all personal communication, other scales, or different organizational cultures. The use of 90-day window on textual data missed longer-term behavioral patterns and richer information about the individual. Hence, we would recommend carefully building upon findings of our work and also encourage future work to explore this with more general populations, other data sources, and more modalities of data.

\paragraph{\textbf{Model constraints.}}
  The choice of prompting instead of more advanced techniques such as fine-tuning limited our ability to correct systematic biases; this choice was largely driven by the time and cost of running the study with 20 participants. Also, although we iterated our pipeline extensively before running the study, some fundamental problems with LLMs could not be avoided via better prompts alone. For example, we observed a few cases of hallucinated outputs as the LLM attributed other speakers' words to participants in meeting transcripts, and occasionally added sentences that didn't exist in the data. However, these were mostly surface-level additions, not fabrication of new data that would affect assessments.

\paragraph{\textbf{Longitudinal evaluation upon deployment.}}
 Our 2-hour sessions captured only initial reactions. The 10 workplace scenarios were hypothetical, but future work could explore if participants' preferences hold when AI responses have real consequences for their relationships and reputation. Participants also did not experience extended use in practice. Future work should deploy this in real life and test whether the auditing improvements persist over time or degrade as contexts change.

%% file: sections/10-conclusion.tex
\section{Conclusion}
We presented \system, an approach for building social profiles from communication data.
\system constructs social style profiles that make inferences from observed social interactions to assess validated psychometric constructs.
In a 20-participant case study with workplace communication scenarios, \system produced initial profiles that participants largely recognized as capturing their personal communication patterns.
Our findings also revealed systematic biases on a few dimensions.
The profile review phase supported calibration, moving participants toward more aligned profiles.
The scenario-based evaluation shows that \system-generated profiled responses were preferred over generic and self-report baselines on aggregate, with preferences varying across individuals and scenario types.
We characterize sources of misalignment and surface challenges around data coverage, boundaries of self-representation, and the desirable degree of representation fidelity.
Our findings from one workplace setting provide an empirical starting point; the natural next steps are to vary the psychometric instrument, broaden the data sources, and study how profiles hold up over time and across organizational cultures.

%% file: sections/appendix.tex
\section*{Appendix}

\section{Evidence Extraction Schema}
\label{app:evidence-schema}

For each behavioral evidence instance extracted during profiling (Section~3.1), the pipeline produces a structured context summary and a short conversational excerpt. Placeholders are shown in \texttt{code} font.

\noindent\textbf{Context (per example)}
\begin{lstlisting}
{
  "situational_background": "What was happening (meeting purpose, topic, timing)",
  "social_dynamics": "Who was involved and their role relative to {user_name}",
  "communication_setting": "1-on-1 vs group; formal vs informal; planned vs spontaneous; stakes",
  "behavioral_analysis": "How context shaped {user_name}'s {facet_name}; would it differ elsewhere?",
  "contextual_significance": "Why this demonstrates {facet_name} given the situation and dynamics"
}
\end{lstlisting}

\noindent\textbf{Conversational Excerpt (2--5 turns)}
\begin{lstlisting}
[
  {"speaker": "User", "message": "Target user's message ..."},
  {"speaker": "OtherPartyName", "message": "Response ..."},
  {"speaker": "User", "message": "Follow-up ..."}
]
\end{lstlisting}

\section{Communication Styles Inventory (CSI) - Complete Scale}
\label{app:csi-full}

The Communication Styles Inventory (CSI)~\cite{devries2013communication} consists of 92 items organized into 6 dimensions and 23 facets. All items are rated on a 5-point Likert scale (1 = strongly disagree, 5 = strongly agree). Items marked with (R) are reverse-coded.

\subsubsection*{Dimension 1: Expressiveness (X)}

\subsubsection*{Talkativeness}
\begin{enumerate}
\item[1.] I always have a lot to say.
\item[25.] I have a hard time keeping myself silent when around other people.
\item[49.] I am never the one who breaks a silence by starting to talk. (R)
\item[73.] I like to talk a lot.
\end{enumerate}

\subsubsection*{Conversational Dominance}
\begin{enumerate}
\item[7.] I often take the lead in a conversation.
\item[31.] Most of the time, other people determine what the discussion is about, not me. (R)
\item[55.] I often determine which topics are talked about during a conversation.
\item[79.] I often determine the direction of a conversation.
\end{enumerate}

\subsubsection*{Humor}
\begin{enumerate}
\item[13.] Because of my humor, I'm often the centre of attention among a group of people.
\item[37.] I have a hard time being humorous in a group. (R)
\item[61.] My jokes always draw a lot of attention.
\item[85.] I often manage to make others burst out laughing.
\end{enumerate}

\subsubsection*{Informality}
\begin{enumerate}
\item[19.] I communicate with others in a distant manner. (R)
\item[43.] I behave somewhat formally when I meet someone. (R)
\item[67.] I address others in a very casual way.
\item[91.] I come across as somewhat stiff when dealing with people. (R)
\end{enumerate}

\subsection*{Dimension 2: Preciseness (P)}

\subsubsection*{Structuredness}
\begin{enumerate}
\item[2.] When I tell a story, the different parts are always clearly related to each other.
\item[26.] I sometimes find it hard to tell a story in an organized way. (R)
\item[50.] I always express a clear chain of thoughts when I argue a point.
\item[74.] My stories always contain a logical structure.
\end{enumerate}

\subsubsection*{Thoughtfulness}
\begin{enumerate}
\item[8.] I think carefully before I say something.
\item[32.] I weigh my answers carefully.
\item[56.] The statements I make are not always well thought out. (R)
\item[80.] I choose my words with care.
\end{enumerate}

\subsubsection*{Substantiveness}
\begin{enumerate}
\item[14.] Conversations with me always involve some important topic.
\item[38.] You won't hear me jabbering about superficial or shallow matters.
\item[62.] I am someone who can often talk about trivial things. (R)
\item[86.] I rarely if ever just chatter away about something.
\end{enumerate}

\subsubsection*{Conciseness}
\begin{enumerate}
\item[20.] I don't need a lot of words to get my message across.
\item[44.] Most of the time, I only need a few words to explain something.
\item[68.] I am somewhat long-winded when I need to explain something. (R)
\item[92.] With a few words I can usually clarify my point to everybody.
\end{enumerate}

\subsection*{Dimension 3: Verbal Aggressiveness (VA)}

\subsubsection*{Angriness}
\begin{enumerate}
\item[3.] If something displeases me, I sometimes explode with anger.
\item[27.] Even when I'm angry, I won't take it out on someone else. (R)
\item[51.] I tend to snap at people when I get annoyed.
\item[75.] I can sometimes react somewhat irritably to people.
\end{enumerate}

\subsubsection*{Authoritarianism}
\begin{enumerate}
\item[9.] I am not very likely to tell someone what they should do. (R)
\item[33.] I sometimes insist that others do what I say.
\item[57.] I expect people to obey when I ask them to do something.
\item[81.] When I feel others should do something for me, I ask for it in a demanding tone of voice.
\end{enumerate}

\subsubsection*{Derogatoriness}
\begin{enumerate}
\item[15.] I never make fun of anyone in a way that might hurt their feelings. (R)
\item[39.] I have at times made people look like fools.
\item[63.] I have been known to be able to laugh at people in their face.
\item[87.] I have humiliated someone in front of a crowd.
\end{enumerate}

\subsubsection*{Nonsupportiveness}
\begin{enumerate}
\item[21.] I can listen well. (R)
\item[45.] I always show a lot of understanding for other people's problems. (R)
\item[69.] I always take time for someone if they want to talk to me. (R)
\item[93.] I always treat people with a lot of respect. (R)
\end{enumerate}

\subsection*{Dimension 4: Questioningness (Q)}

\subsubsection*{Unconventionality}
\begin{enumerate}
\item[4.] I sometimes toss bizarre ideas into a group discussion.
\item[28.] I often say unexpected things.
\item[52.] In discussions, I often put forward unusual points of view.
\item[76.] In conversations, I often toy with some very wild ideas.
\end{enumerate}

\subsubsection*{Philosophicalness}
\begin{enumerate}
\item[10.] I never enter into discussions about the future of the human race. (R)
\item[34.] I like to talk with others about the deeper aspects of our existence.
\item[58.] I never engage in so-called philosophical conversations. (R)
\item[82.] I regularly have discussions with people about the meaning of life.
\end{enumerate}

\subsubsection*{Inquisitiveness}
\begin{enumerate}
\item[16.] During a conversation, I always try to find out about the background of somebody's opinion.
\item[40.] I don't bother asking a lot of questions just to find out why people feel the way they do about something. (R)
\item[64.] I ask a lot of questions to uncover someone's motives.
\item[88.] I always ask how people arrive at their conclusions.
\end{enumerate}

\subsubsection*{Argumentativeness}
\begin{enumerate}
\item[22.] To stimulate discussion, I sometimes express a view different from that of my conversation partner.
\item[46.] I like to provoke others by making bold statements.
\item[70.] I try to find out what people think about a topic by getting them to debate with me about it.
\item[94.] By making controversial statements, I often force people to express a clear opinion.
\end{enumerate}

\subsection*{Dimension 5: Emotionality (E)}

\subsubsection*{Sentimentality}
\begin{enumerate}
\item[5.] When I see others cry, I have difficulty holding back my tears.
\item[29.] During a conversation, I am not easily overcome by emotions. (R)
\item[53.] When describing my memories, I sometimes get visibly emotional.
\item[77.] People can tell that I am emotionally touched by some topics of conversation.
\end{enumerate}

\subsubsection*{Worrisomeness}
\begin{enumerate}
\item[11.] When I'm worried about something, I find it hard to talk about anything else.
\item[35.] I tend to talk about my concerns a lot.
\item[59.] People can tell when I feel anxious.
\item[83.] When I worry, everybody notices.
\end{enumerate}

\subsubsection*{Tension}
\begin{enumerate}
\item[17.] Because of stress, I am sometimes unable to express myself properly.
\item[41.] I can be visibly tense during a conversation.
\item[65.] I am able to address a large group of people very calmly. (R)
\item[89.] I find it hard to talk in a relaxed manner when what I have to say is valued highly.
\end{enumerate}

\subsubsection*{Defensiveness}
\begin{enumerate}
\item[23.] The comments of others have a noticeable effect on me.
\item[47.] Nasty remarks from other people do not bother me too much. (R)
\item[71.] When people criticize me, I am visibly hurt.
\item[95.] I am not always able to cope easily with critical remarks.
\end{enumerate}

\subsection*{Dimension 6: Impression Manipulativeness (IM)}

\subsubsection*{Ingratiation}
\begin{enumerate}
\item[6.] I sometimes praise somebody at great length, without being really genuine, in order to make them like me.
\item[30.] In discussions I sometimes express an opinion I do not support in order to make a good impression.
\item[54.] Sometimes I use flattery to get someone in a favorable mood.
\item[78.] To be considered likeable, I sometimes say things my conversation partner likes to hear.
\end{enumerate}

\subsubsection*{Charm}
\begin{enumerate}
\item[12.] I sometimes use my charm to get something done.
\item[36.] I sometimes flirt a little bit to win somebody over.
\item[60.] I would not use my appearance to make people do things for me. (R)
\item[84.] I sometimes put on a very seductive voice when I want something.
\end{enumerate}

\subsubsection*{Concealingness}
\begin{enumerate}
\item[24.] I sometimes conceal information to make me look better.
\item[48.] I sometimes "forget" to tell something when this is more convenient for me.
\item[72.] I tell people the whole story, even when this is probably not good for me. (R)
\item[96.] Even if I would benefit from withholding information from someone, I would find it hard to do so. (R)
\end{enumerate}

\subsection*{Scoring Instructions}
\begin{enumerate}
\item Reverse code all items marked with (R) using the formula: 1→5, 2→4, 3→3, 4→2, 5→1
\item Calculate facet scores by averaging the 4 items within each facet
\item Calculate dimension scores by averaging the facet scores within each dimension
\item Note: The original CSI had a fourth facet (Inscrutableness) under Impression Manipulativeness that was removed due to poor psychometric properties hence the highest number here is 96, but the total number of items is 92.
\end{enumerate}

\section{Scenario templates used in study}
\label{appendix:scenarios}

\begin{itemize}
    \item S1: Weekly team check-in opening - Target dimension: Expressiveness
    \item S2: Explaining your work to interested colleague - Target dimension: Preciseness
    \item S3: Colleague questions your approach - Target dimension: Verbal Aggressiveness
    \item S4: New initiative announcement - Target dimension: Questioningness
    \item S5: Last-minute schedule change - Target dimension: Emotionality
    \item S6: Informal catch-up with influential colleague - Target dimension: Impression Manipulativeness
    \item S7: Sharing project update at standup - Target dimension: Expressiveness × Preciseness
    \item S8: Unexpected process change notification - Target dimension: Questioningness × Emotionality
    \item S9: Credit attribution discussion - Target dimension: Verbal Aggressiveness × Impression Manipulativeness
    \item S10: Team celebration planning - Target dimension: Emotionality × Expressiveness
\end{itemize}

\begin{table*}[ht]
\centering
\caption{APRACE Attributes by Scenario}
\label{tab:aprace-attributes}
\scriptsize
\begin{tabular}{lccccccccccc}
\toprule
Scenario & Hierarchy & Familiarity & Purpose & Mode & Stakes & Formality & Timing & Audience & Emotional & Motivation & Desired \\
 & & & & & & & & & State & & Outcome \\
\midrule
S1 & Peer & Close & Info. Sharing & Video Call & Low & Informal & Routine & Small Group & Neutral & Social & Positive Res. \\
S2 & Peer & Distant & Info. Sharing & Face-to-Face & Low & Informal & Routine & Private & Neutral & Achievement & Neutral Comp. \\
S3 & Peer & Close & Decision Making & Chat & Medium & Informal & Routine & Private & Neutral & Achievement & Positive Res. \\
S4 & Manager & Close & Planning & Face-to-Face & Medium & Formal & Scheduled & Small Group & Confident & Autonomy & Neutral Comp. \\
S5 & Collaborator & Distant & Problem Solving & Video Call & Medium & Formal & Urgent & Private & Stressed & Security & Conflict Avoid. \\
S6 & Collaborator & New & Social & Face-to-Face & Low & Informal & Routine & Private & Neutral & Social & Positive Res. \\
S7 & Peer & Close & Info. Sharing & Video Call & Low & Informal & Routine & Small Group & Confident & Achievement & Positive Res. \\
S8 & Subordinate & Close & Info. Sharing & Chat & Medium & Informal & Routine & Private & Neutral & Autonomy & Neutral Comp. \\
S9 & Peer & Distant & Problem Solving & Face-to-Face & Medium & Informal & Scheduled & Private & Frustrated & Achievement & Positive Res. \\
S10 & Peer & Close & Social & Face-to-Face & Low & Informal & Routine & Small Group & Confident & Social & Positive Res. \\
\bottomrule
\end{tabular}
\end{table*}

\section{Data Collected}
\label{sec:appendix-data-collected}
\textbf{Definition.} We distinguish \emph{raw} data (participant-generated or system-logged inputs as collected, after de-identification) from \emph{derived} data (outputs produced by models or researchers).

\subsection{Raw data from participants.}
\begin{enumerate}
  \item \textbf{Psychometric self-report (CSI).} 92 items across 6 dimensions/23 facets, Likert 1–5, plus optional free-text clarifications ($N{=}20$; $1{,}840$ item–person pairs).
  \item \textbf{Scenario evaluations.} Within-subject triad comparisons for each scenario: rank-order (1st/2nd/3rd) and Likert appropriateness ratings for \emph{Profiled}, \emph{Generic}, and \emph{Self-Report} responses (600 total evaluations), with optional short rationales.
  \item \textbf{Video recordings of audit interactions.} Think aloud rationales and decisions for per-facet review actions during auditing.
\end{enumerate}

\subsection{Derived data (model-produced).}
\begin{enumerate}
  \item \textbf{LLM-inferred initial profiles.} Item- and facet-level predictions from the communication corpus mapped to CSI constructs.
  \item \textbf{Generated scenario responses.} Model outputs for each condition (Profiled, Generic, Self-Report) per scenario template; canonicalized text with metadata.
\end{enumerate}

\noindent\textbf{Note.} All raw text was never sent to the researchers; derived artifacts were produced by our pipeline installed on participants laptop during the study and sent to researchers after review.

\newpage
\section{Figures}
\begin{figure*}[h!]
  \centering
  \includegraphics[width=\textwidth]{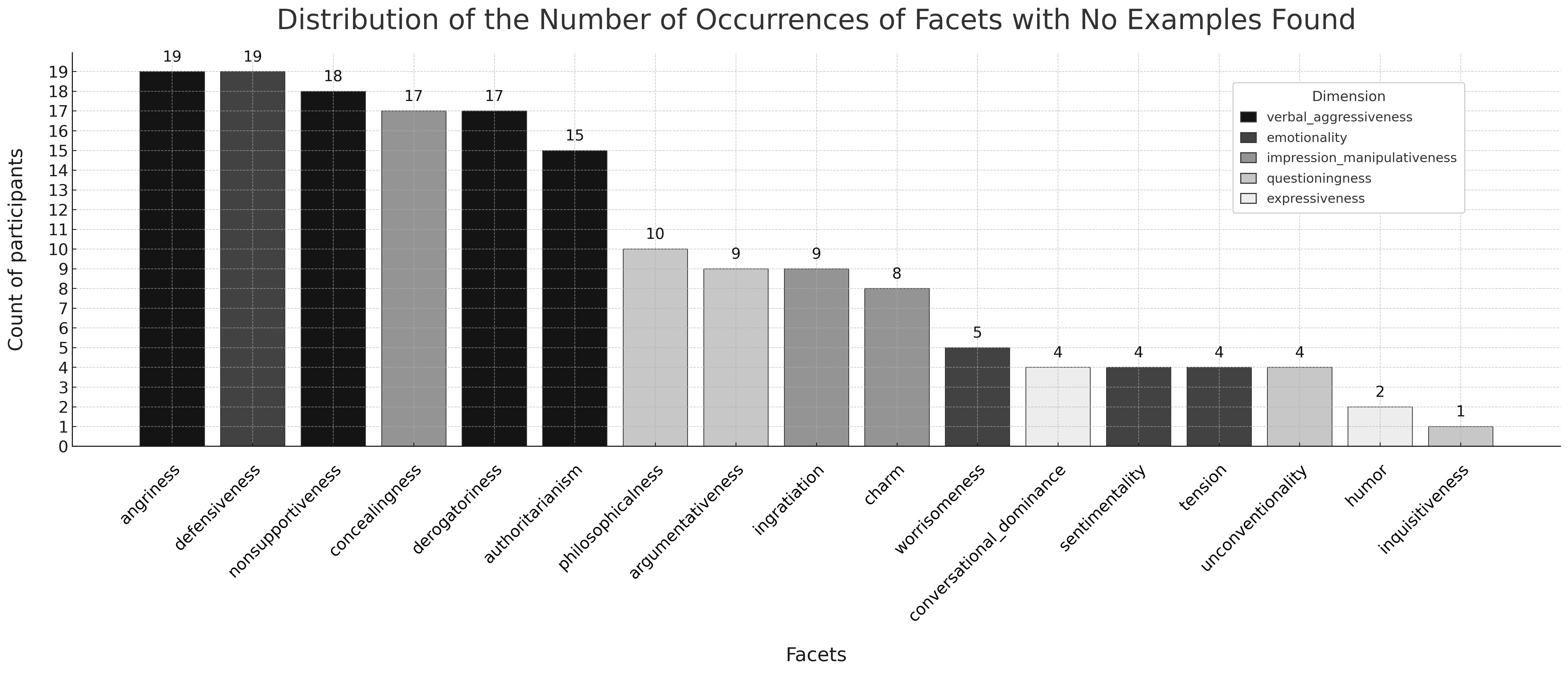} 
  \caption{Distribution of occurrences of facets with no examples found in data across 20 participants. Each facet is assessed once, so the maximum number of occurrences would be 20 per facet. This is evident of what traits tend to be lack of behavioral evidence from workplace data.}
  \label{fig:no_examples}
\end{figure*}

%% file: references.bib
@misc{cheng2024gems,
      title={GEMS: Generative Expert Metric System through Iterative Prompt Priming}, 
      author={Ti-Chung Cheng and Carmen Badea and Christian Bird and Thomas Zimmermann and Robert DeLine and Nicole Forsgren and Denae Ford},
      year={2024},
      eprint={2410.00880},
      archivePrefix={arXiv},
      primaryClass={cs.SE},
      url={https://arxiv.org/abs/2410.00880}, 
}

@ARTICLE{hoppler2022aprace,
AUTHOR={Hoppler, Sarah Susanna  and Segerer, Robin  and Nikitin, Jana },
TITLE={The Six Components of Social Interactions: Actor, Partner, Relation, Activities, Context, and Evaluation},
JOURNAL={Frontiers in Psychology},
VOLUME={Volume 12 - 2021},
YEAR={2022},
URL={https://www.frontiersin.org/journals/psychology/articles/10.3389/fpsyg.2021.743074},
DOI={10.3389/fpsyg.2021.743074},
ISSN={1664-1078},
}

@online{patel2024zoom,
  author    = {Nilay Patel},
  title     = {The CEO of Zoom wants AI clones in meetings},
  year      = {2024},
  url       = {https://www.theverge.com/2024/6/3/24168733/zoom-ceo-ai-clones-digital-twins-videoconferencing-decoder-interview},
  note      = {Accessed: 2024-09-04},
  journal   = {The Verge}
}

@online{characterai2025,
  title        = {Character.{AI}},
  year         = {2025},
  note         = {Founded 2021; public beta launched September 16, 2022; founders Noam Shazeer and Daniel de Freitas},
  url          = {https://character.ai/},
  organization = {Character.AI}
}

@inproceedings{park2023generative,
  title={Generative agents: Interactive simulacra of human behavior},
  author={Park, Joon Sung and O'Brien, Joseph and Cai, Carrie Jun and Morris, Meredith Ringel and Liang, Percy and Bernstein, Michael S},
  booktitle={Proceedings of the 36th annual acm symposium on user interface software and technology},
  pages={1--22},
  year={2023}
}

@inproceedings{cheng2025conversational,
  title={Conversational Agents on Your Behalf: Opportunities and Challenges of Shared Autonomy in Voice Communication for Multitasking},
  author={Cheng, Yi Fei and Shirado, Hirokazu and Kasahara, Shunichi},
  booktitle={Proceedings of the 2025 CHI Conference on Human Factors in Computing Systems},
  pages={1--18},
  year={2025}
}

@article{lee2023speculating,
  title={Speculating on risks of AI clones to selfhood and relationships: Doppelganger-phobia, identity fragmentation, and living memories},
  author={Lee, Patrick Yung Kang and Ma, Ning F and Kim, Ig-Jae and Yoon, Dongwook},
  journal={Proceedings of the ACM on Human-computer Interaction},
  volume={7},
  number={CSCW1},
  pages={1--28},
  year={2023},
  publisher={ACM New York, NY, USA}
}

@inproceedings{sharma2021towards,
  title={Towards facilitating empathic conversations in online mental health support: A reinforcement learning approach},
  author={Sharma, Ashish and Lin, Inna W and Miner, Adam S and Atkins, David C and Althoff, Tim},
  booktitle={Proceedings of the web conference 2021},
  pages={194--205},
  year={2021}
}

@article{ziems2022inducing,
  title={Inducing positive perspectives with text reframing},
  author={Ziems, Caleb and Li, Minzhi and Zhang, Anthony and Yang, Diyi},
  journal={arXiv preprint arXiv:2204.02952},
  year={2022}
}

@article{hancock2020ai,
  title={AI-mediated communication: Definition, research agenda, and ethical considerations},
  author={Hancock, Jeffrey T and Naaman, Mor and Levy, Karen},
  journal={Journal of Computer-Mediated Communication},
  volume={25},
  number={1},
  pages={89--100},
  year={2020},
  publisher={Oxford University Press}
}

@article{pavlick2016empirical,
  title={An empirical analysis of formality in online communication},
  author={Pavlick, Ellie and Tetreault, Joel},
  journal={Transactions of the association for computational linguistics},
  volume={4},
  pages={61--74},
  year={2016},
  publisher={MIT Press One Rogers Street, Cambridge, MA 02142-1209, USA journals-info~…}
}

@inproceedings{ma2017self,
  title={Self-disclosure and perceived trustworthiness of Airbnb host profiles},
  author={Ma, Xiao and Hancock, Jeffrey T and Lim Mingjie, Kenneth and Naaman, Mor},
  booktitle={Proceedings of the 2017 ACM conference on computer supported cooperative work and social computing},
  pages={2397--2409},
  year={2017}
}

@article{kim2019love,
  title={Love in lyrics: An exploration of supporting textual manifestation of affection in social messaging},
  author={Kim, Taewook and Lee, Jung Soo and Peng, Zhenhui and Ma, Xiaojuan},
  journal={Proceedings of the ACM on Human-Computer Interaction},
  volume={3},
  number={CSCW},
  pages={1--27},
  year={2019},
  publisher={ACM New York, NY, USA}
}

@article{argyle2023leveraging,
  title={Leveraging AI for democratic discourse: Chat interventions can improve online political conversations at scale},
  author={Argyle, Lisa P and Bail, Christopher A and Busby, Ethan C and Gubler, Joshua R and Howe, Thomas and Rytting, Christopher and Sorensen, Taylor and Wingate, David},
  journal={Proceedings of the National Academy of Sciences},
  volume={120},
  number={41},
  pages={e2311627120},
  year={2023},
  publisher={National Academy of Sciences}
}

@article{leong2024dittos,
  title={Dittos: Personalized, embodied agents that participate in meetings when you are unavailable},
  author={Leong, Joanne and Tang, John and Cutrell, Edward and Junuzovic, Sasa and Baribault, Gregory Paul and Inkpen, Kori},
  journal={Proceedings of the ACM on Human-Computer Interaction},
  volume={8},
  number={CSCW2},
  pages={1--28},
  year={2024},
  publisher={ACM New York, NY, USA}
}

@inproceedings{huang2025mirror,
  title={Mirror to Companion: Exploring Roles, Values, and Risks of AI Self-Clones through Story Completion},
  author={Huang, Jessica and Kim, Ig-Jae and Yoon, Dongwook},
  booktitle={Proceedings of the 2025 CHI Conference on Human Factors in Computing Systems},
  pages={1--15},
  year={2025}
}

@inproceedings{hwang2024whose,
  title={In whose voice?: examining AI agent representation of people in social interaction through generative speech},
  author={Hwang, Angel Hsing-Chi and Siy, John Oliver and Shelby, Renee and Lentz, Alison},
  booktitle={Proceedings of the 2024 ACM Designing Interactive Systems Conference},
  pages={224--245},
  year={2024}
}

@article{zakazov2024assessing,
  title={Assessing Social Alignment: Do Personality-Prompted Large Language Models Behave Like Humans?},
  author={Zakazov, Ivan and Boronski, Mikolaj and Drudi, Lorenzo and West, Robert},
  journal={arXiv preprint arXiv:2412.16772},
  year={2024}
}

@article{park2024generative,
  title={Generative agent simulations of 1,000 people},
  author={Park, Joon Sung and Zou, Carolyn Q and Shaw, Aaron and Hill, Benjamin Mako and Cai, Carrie and Morris, Meredith Ringel and Willer, Robb and Liang, Percy and Bernstein, Michael S},
  journal={arXiv preprint arXiv:2411.10109},
  year={2024}
}

@article{shaikh2025creating,
  title={Creating General User Models from Computer Use},
  author={Shaikh, Omar and Sapkota, Shardul and Rizvi, Shan and Horvitz, Eric and Park, Joon Sung and Yang, Diyi and Bernstein, Michael S},
  journal={arXiv preprint arXiv:2505.10831},
  year={2025}
}

@inproceedings{jakesch2019ai,
  title={AI-mediated communication: How the perception that profile text was written by AI affects trustworthiness},
  author={Jakesch, Maurice and French, Megan and Ma, Xiao and Hancock, Jeffrey T and Naaman, Mor},
  booktitle={Proceedings of the 2019 CHI conference on human factors in computing systems},
  pages={1--13},
  year={2019}
}

@inproceedings{mcilroy2022mimetic,
  title={Mimetic models: Ethical implications of ai that acts like you},
  author={McIlroy-Young, Reid and Kleinberg, Jon and Sen, Siddhartha and Barocas, Solon and Anderson, Ashton},
  booktitle={Proceedings of the 2022 AAAI/ACM Conference on AI, Ethics, and Society},
  pages={479--490},
  year={2022}
}

@inproceedings{hohenstein2018ai,
  title={AI-supported messaging: An investigation of human-human text conversation with AI support},
  author={Hohenstein, Jess and Jung, Malte},
  booktitle={Extended abstracts of the 2018 CHI conference on human factors in computing systems},
  pages={1--6},
  year={2018}
}

@inproceedings{zhou2017adapting,
  title={Adapting a persuasive conversational agent for the Chinese culture},
  author={Zhou, Shuo and Zhang, Zhe and Bickmore, Timothy},
  booktitle={2017 international conference on culture and computing (culture and computing)},
  pages={89--96},
  year={2017},
  organization={IEEE}
}

@inproceedings{nielsen2014personas,
  title={Personas is applicable: a study on the use of personas in Denmark},
  author={Nielsen, Lene and Storgaard Hansen, Kira},
  booktitle={Proceedings of the SIGCHI Conference on Human Factors in Computing Systems},
  pages={1665--1674},
  year={2014}
}

@article{cronbach1955construct,
  title={Construct validity in psychological tests.},
  author={Cronbach, Lee J and Meehl, Paul E},
  journal={Psychological bulletin},
  volume={52},
  number={4},
  pages={281},
  year={1955},
  publisher={American Psychological Association}
}

@book{devellis2021scale,
  title={Scale development: Theory and applications},
  author={DeVellis, Robert F and Thorpe, Carolyn T},
  year={2021},
  publisher={Sage publications}
}

@article{goldberg1993structure,
  title={The structure of phenotypic personality traits.},
  author={Goldberg, Lewis R},
  journal={American psychologist},
  volume={48},
  number={1},
  pages={26},
  year={1993},
  publisher={American Psychological Association}
}

@article{costa2008revised,
  title={The revised neo personality inventory (neo-pi-r)},
  author={Costa, Paul T and McCrae, Robert R},
  journal={The SAGE handbook of personality theory and assessment},
  volume={2},
  number={2},
  pages={179--198},
  year={2008}
}

@article{devries2013communication,
  title={The communication styles inventory (CSI) a six-dimensional behavioral model of communication styles and its relation with personality},
  author={De Vries, Reinout E and Bakker-Pieper, Angelique and Konings, Femke E and Schouten, Barbara},
  journal={Communication Research},
  volume={40},
  number={4},
  pages={506--532},
  year={2013},
  publisher={Sage Publications Sage CA: Los Angeles, CA}
}

@article{kosinski2013private,
  title={Private traits and attributes are predictable from digital records of human behavior},
  author={Kosinski, Michal and Stillwell, David and Graepel, Thore},
  journal={Proceedings of the national academy of sciences},
  volume={110},
  number={15},
  pages={5802--5805},
  year={2013},
  publisher={National Academy of Sciences}
}

@article{youyou2015computer,
  title={Computer-based personality judgments are more accurate than those made by humans},
  author={Youyou, Wu and Kosinski, Michal and Stillwell, David},
  journal={Proceedings of the National Academy of Sciences},
  volume={112},
  number={4},
  pages={1036--1040},
  year={2015},
  publisher={National Academy of Sciences}
}

@article{kosinski2015facebook,
  title={Facebook as a research tool for the social sciences: Opportunities, challenges, ethical considerations, and practical guidelines.},
  author={Kosinski, Michal and Matz, Sandra C and Gosling, Samuel D and Popov, Vesselin and Stillwell, David},
  journal={American psychologist},
  volume={70},
  number={6},
  pages={543},
  year={2015},
  publisher={American Psychological Association}
}

@inproceedings{baik2025adapting,
  title={Adapting Communication Styles in Health Chatbot using Large Language Models to Support Family Caregivers from Multicultural Backgrounds},
  author={Baik, Rebekah Lee and Lee, Stephanie and Xie, Serena Jinchen and Liao, Wang and Hwang, Elina H and Yuwen, Weichao},
  booktitle={Proceedings of the Extended Abstracts of the CHI Conference on Human Factors in Computing Systems},
  pages={1--8},
  year={2025}
}

@article{bakker2013incremental,
  title={The incremental validity of communication styles over personality traits for leader outcomes},
  author={Bakker-Pieper, Angelique and de Vries, Reinout E},
  journal={Human Performance},
  volume={26},
  number={1},
  pages={1--19},
  year={2013},
  publisher={Taylor \& Francis}
}

@article{kluger1996effects,
  title={The effects of feedback interventions on performance: a historical review, a meta-analysis, and a preliminary feedback intervention theory.},
  author={Kluger, Avraham N and DeNisi, Angelo},
  journal={Psychological bulletin},
  volume={119},
  number={2},
  pages={254},
  year={1996},
  publisher={American Psychological Association}
}

@article{diotaiuti2020psychometric,
  title={Psychometric properties and a preliminary validation study of the Italian brief version of the communication styles inventory (CSI-B/I)},
  author={Diotaiuti, Pierluigi and Valente, Giuseppe and Mancone, Stefania and Grambone, Angela},
  journal={Frontiers in Psychology},
  volume={11},
  pages={1421},
  year={2020},
  publisher={Frontiers Media SA}
}

@article{gebru2021datasheets,
  title={Datasheets for datasets},
  author={Gebru, Timnit and Morgenstern, Jamie and Vecchione, Briana and Vaughan, Jennifer Wortman and Wallach, Hanna and Iii, Hal Daum{\'e} and Crawford, Kate},
  journal={Communications of the ACM},
  volume={64},
  number={12},
  pages={86--92},
  year={2021},
  publisher={ACM New York, NY, USA}
}

@article{lee2004trust,
  title={Trust in automation: Designing for appropriate reliance},
  author={Lee, John D and See, Katrina A},
  journal={Human factors},
  volume={46},
  number={1},
  pages={50--80},
  year={2004},
  publisher={SAGE Publications Sage UK: London, England}
}

@inproceedings{mitchell2019model,
  title={Model cards for model reporting},
  author={Mitchell, Margaret and Wu, Simone and Zaldivar, Andrew and Barnes, Parker and Vasserman, Lucy and Hutchinson, Ben and Spitzer, Elena and Raji, Inioluwa Deborah and Gebru, Timnit},
  booktitle={Proceedings of the conference on fairness, accountability, and transparency},
  pages={220--229},
  year={2019}
}

@inproceedings{raji2020closing,
  title={Closing the AI accountability gap: Defining an end-to-end framework for internal algorithmic auditing},
  author={Raji, Inioluwa Deborah and Smart, Andrew and White, Rebecca N and Mitchell, Margaret and Gebru, Timnit and Hutchinson, Ben and Smith-Loud, Jamila and Theron, Daniel and Barnes, Parker},
  booktitle={Proceedings of the 2020 conference on fairness, accountability, and transparency},
  pages={33--44},
  year={2020}
}

@article{bell1984language,
  title={Language style as audience design},
  author={Bell, Allan},
  journal={Language in society},
  volume={13},
  number={2},
  pages={145--204},
  year={1984},
  publisher={Cambridge University Press}
}

@misc{brown1987politeness,
  title={Politeness: Some universals in language usage},
  author={Brown, Penelope},
  year={1987},
  publisher={Cambridge university press}
}

@article{giles1991accommodation,
  title={Accommodation theory: Communication, context, and consequence},
  author={Giles, Howard and Coupland, Nikolas and Coupland, Justine},
  journal={Contexts of accommodation: Developments in applied sociolinguistics},
  volume={1},
  pages={1--68},
  year={1991},
  publisher={Cambridge}
}

@article{goffman1959presentation,
  title={The presentation of self in everyday life, Double Day Anchor},
  author={Goffman, Erving},
  journal={Garden City, NY},
  year={1959}
}

@article{chi1994eliciting,
  title={Eliciting self-explanations improves understanding},
  author={Chi, Michelene TH and De Leeuw, Nicholas and Chiu, Mei-Hung and LaVancher, Christian},
  journal={Cognitive science},
  volume={18},
  number={3},
  pages={439--477},
  year={1994},
  publisher={Elsevier}
}

@article{mabe1982validity,
  title={Validity of self-evaluation of ability: A review and meta-analysis.},
  author={Mabe, Paul A and West, Stephen G},
  journal={Journal of applied Psychology},
  volume={67},
  number={3},
  pages={280},
  year={1982},
  publisher={American Psychological Association}
}

@article{kruger1999unskilled,
  title={Unskilled and unaware of it: how difficulties in recognizing one's own incompetence lead to inflated self-assessments.},
  author={Kruger, Justin and Dunning, David},
  journal={Journal of personality and social psychology},
  volume={77},
  number={6},
  pages={1121},
  year={1999},
  publisher={American Psychological Association}
}

@article{vazire2010knows,
  title={Who knows what about a person? The self--other knowledge asymmetry (SOKA) model.},
  author={Vazire, Simine},
  journal={Journal of personality and social psychology},
  volume={98},
  number={2},
  pages={281},
  year={2010},
  publisher={American Psychological Association}
}

@article{connelly2010other,
  title={An other perspective on personality: meta-analytic integration of observers' accuracy and predictive validity.},
  author={Connelly, Brian S and Ones, Deniz S},
  journal={Psychological bulletin},
  volume={136},
  number={6},
  pages={1092},
  year={2010},
  publisher={American Psychological Association}
}

@article{ouyang2022training,
  title={Training language models to follow instructions with human feedback},
  author={Ouyang, Long and Wu, Jeffrey and Jiang, Xu and Almeida, Diogo and Wainwright, Carroll and Mishkin, Pamela and Zhang, Chong and Agarwal, Sandhini and Slama, Katarina and Ray, Alex and others},
  journal={Advances in neural information processing systems},
  volume={35},
  pages={27730--27744},
  year={2022}
}

@article{rafailov2023direct,
  title={Direct preference optimization: Your language model is secretly a reward model},
  author={Rafailov, Rafael and Sharma, Archit and Mitchell, Eric and Manning, Christopher D and Ermon, Stefano and Finn, Chelsea},
  journal={Advances in neural information processing systems},
  volume={36},
  year={2023}
}

@inproceedings{poddar2024personalizing,
  title={Personalizing Reinforcement Learning from Human Feedback with Variational Preference Learning},
  author={Poddar, Sriyash and Wan, Yanming and Ivison, Hamish and Choudhury, Siddharth and Jaques, Natasha},
  booktitle={NeurIPS 2024 Workshop on Pluralistic Alignment},
  year={2024}
}

@inproceedings{chakraborty2024maxmin,
  title={MaxMin-RLHF: Alignment with diverse human preferences},
  author={Chakraborty, Souradip and Qin, Jiahao and Garcelon, Evrard and Lazaric, Alessandro and Pirotta, Matteo and Zanette, Andrea},
  booktitle={Proceedings of the 41st International Conference on Machine Learning},
  year={2024}
}
